\documentclass[12pt]{article}

\usepackage{amsmath}
\usepackage{amssymb}
\usepackage{fancyheadings}

\numberwithin{equation}{section}

\usepackage{ifthen}

\newcommand{\REC}{ {\rm REC } }

\newcommand{\rec}{ {\rm rec } }

\newcommand{\DTIME} { {\rm DTIME}}

\newcommand{\set}[2]{\left\{ {#1} \: \Big| \: {#2} \right\} }
\newcommand{\strings}{\left\{ 0, 1 \right\}^\ast}

\newcommand{\func}[3]{ {#1} : {#2} \rightarrow {#3} }
\newcommand{\N}{{\mathbb N}}
\newcommand{\Q}{{\mathbb Q}}
\newcommand{\R}{{\mathbb R}}

\newcommand{\one}{{\mathbf 1}}

\newcommand{\bvec}{{\overrightarrow{\beta}}}
\newcommand{\psqsubset}{\mbox{$\:^{\sqsubset}_{\neq}\:$}}
\newcommand{\bval}[1]{[\![ #1 ]\!]}

\newcommand{\CantorSet}{{\mathbf C}}
\newcommand{\DeltaPrime}{{\Delta^\prime}}
\newcommand{\Xplus}{{X^+}}
\newcommand{\Xminus}{{X^-}}
\newcommand{\Yplus}{{Y^+}}
\newcommand{\Yminus}{{Y^-}}
\newcommand{\dpr}{{d^+_r}}
\newcommand{\dmr}{{d^-_r}}
\newcommand{\distnu}{{ {\cal D}_\nu}}

\newcounter{rctr}[section]
\newtheorem{dummy}{xyzzy}[section]
\newtheorem{xtheorem}[dummy]{\bf Theorem}
\newtheorem{xlemma}[dummy]{\bf Lemma}
\newtheorem{xobservation}[dummy]{\bf Observation}
\newtheorem{xexample}[dummy]{\bf Example}
\newtheorem{xexamples}[dummy]{\bf Examples}
\newtheorem{xremark}[dummy]{\bf Remark}
\newtheorem{xremarks}[dummy]{\bf Remarks}
\newtheorem{xdefinition}[dummy]{\bf Definition}
\newtheorem{xcorollary}[dummy]{\bf Corollary}
\newtheorem{xconstruction}[dummy]{\bf Construction}
\newtheorem{xconjecture}[dummy]{\bf Conjecture}

\newenvironment{theorem}[1][xyzzy]
{
\begin{xtheorem}%
\normalfont%
\ifthenelse{\equal{#1}{xyzzy}}{{\hspace*{-6pt}\bf{.}}}{ (#1). }
}
{
\end{xtheorem}
}

\newenvironment{lemma}[1][xyzzy]
{
\begin{xlemma}%
\normalfont%
\ifthenelse{\equal{#1}{xyzzy}}{{\hspace*{-6pt}\bf{.}}}{ (#1).}
}
{
\end{xlemma}
}

\newenvironment{example}[1][xyzzy]
{
\begin{xexample}%
\normalfont%
\ifthenelse{\equal{#1}{xyzzy}}{{\hspace*{-6pt}\bf{.}}}{ (#1).}
}
{
\end{xexample}
}

\newenvironment{corollary}[1][xyzzy]
{
\begin{xcorollary}%
\normalfont%
\ifthenelse{\equal{#1}{xyzzy}}{{\hspace*{-6pt}\bf{.}}}{ (#1).}
}
{
\end{xcorollary}
}

\newenvironment{theorem*}[1][xyzzy]
{{\noindent {\bf Theorem}%
\ifthenelse{\equal{#1}{xyzzy}}{{\bf .}}{ (#1).}}
}
{
}

\newenvironment{lemma*}[1][xyzzy]
{
{\noindent {\bf Lemma}%
\ifthenelse{\equal{#1}{xyzzy}}{{\bf .}}{ (#1).}}
}
{
}

\newenvironment{observation*}[1][xyzzy]
{{\noindent {\bf Observation}%
\ifthenelse{\equal{#1}{xyzzy}}{{\bf .}}{ (#1).}}
}
{
}

\newenvironment{conjecture*}[1][xyzzy]
{{\noindent {\bf Conjecture}%
\ifthenelse{\equal{#1}{xyzzy}}{{\bf .}}{ (#1).}}
}
{
}

\newenvironment{example*}[1][xyzzy]
{{\noindent {\bf Example}%
\ifthenelse{\equal{#1}{xyzzy}}{{\bf .}}{ (#1).}}
}
{
}

\newenvironment{examples*}[1][xyzzy]
{{\noindent {\bf Examples}%
\ifthenelse{\equal{#1}{xyzzy}}{{\bf .}}{ (#1).}}
}
{
}

\newenvironment{remark*}[1][xyzzy]
{{\noindent {\bf Remark}%
\ifthenelse{\equal{#1}{xyzzy}}{{\bf .}}{ (#1).}}
}
{
}

\newenvironment{remarks*}[1][xyzzy]
{
{\noindent {\bf Remarks}%
\ifthenelse{\equal{#1}{xyzzy}}{{\bf .}}{ (#1).}}
}
{
}

\newenvironment{construction*}[1][xyzzy]
{
{\noindent {\bf Construction}%
\ifthenelse{\equal{#1}{xyzzy}}{{\bf .}}{ (#1).}}
}
{
}

\newenvironment{corollary*}[1][xyzzy]
{
{\noindent {\bf Corollary}%
\ifthenelse{\equal{#1}{xyzzy}}{{\bf .}}{ (#1).}}
}
{
}

\newenvironment{proofof}[1][xyzzy]
{
{\noindent {\bf Proof of #1}}}
{
\hfill $\square$
}

\newenvironment{definition*}[1][xyzzy]
{
{\noindent {\bf Definition}%
\ifthenelse{\equal{#1}{xyzzy}}{{\bf .}}{ (#1).}}
}
{
}

\newenvironment{notation*}[1][xyzzy]
{
{\noindent {\bf Notation}%
\ifthenelse{\equal{#1}{xyzzy}}{{\bf .}}{ (#1).}}
}
{
}

\newcommand{\E}{{\rm E}\ }

\global\def\timestamp{
\TeX~~ started on~~ 
\ifnum \month = 1 Janurary \fi
\ifnum \month = 2 February \fi
\ifnum \month = 3 March \fi
\ifnum \month = 4 April \fi
\ifnum \month = 5 May \fi
\ifnum \month = 6 June \fi
\ifnum \month = 7 July \fi
\ifnum \month = 8 August \fi
\ifnum \month = 9 September \fi
\ifnum \month = 10 October \fi
\ifnum \month = 11 November \fi
\ifnum \month = 12 December \fi
\the\day , \the\year~~~~
\ta=\time
\tb=\ta
\divide\ta by60
\tc=\ta
\multiply\ta by60
\advance\tb by-\ta
\the\tc:\the\tb:00
}

\begin{document}

\pagestyle{empty}
\begin{center}
{\bf \LARGE Resource-Bounded Measure \footnotemark}
\end{center}
\footnotetext{This research was supported in part by National Science
  Foundation Grants 9157382 (with matching funds from Rockwell,
  Microware Systems Corporation, and Amoco Foundation) and 9610461.
  Much of the research was performed during several visits at the
  California Institute of Technology, and much of the writing took
  place during a sabbatical at Cornell University.}

\begin{center}
Jack H. Lutz\\
Department of Computer Science \\
Iowa State University \\
Ames, Iowa  50011 \\
U.S.A.
\end{center}

\begin{abstract}
A general theory of resource-bounded measurability and measure is
developed.  Starting from any feasible probability measure $\nu$ on
the Cantor space $\CantorSet$ (the set of all decision problems) and
any suitable complexity class ${\cal C} \subseteq \CantorSet$, the
theory identifies the subsets of $\CantorSet$ that are
$\nu$-measurable in ${\cal C}$ and assigns measures to these sets,
thereby endowing ${\cal C}$ with internal measure-theoretic
structure.  Classes ${\cal C}$ to which the theory applies include
various exponential time and space complexity classes, the class of
all decidable languages, and the Cantor space $\CantorSet$ itself, on
which the resource-bounded theory is shown to agree with the
classical theory.

The sets that are $\nu$-measurable in ${\cal C}$ are shown to form an
algebra relative to which $\nu$-measure is well-behaved (monotone,
additive, etc.).  This algebra is also shown to be complete (subsets
of measure $0$ sets are measurable) and closed under sufficiently
uniform infinitary unions and intersections, and $\nu$-measure in
${\cal C}$ is shown to have the appropriate additivity and monotone
convergence properties with respect to such infinitary operations.

A generalization of the classical Kolmogorov zero-one law is proven,
showing that when $\nu$ is any feasible coin-toss (i.e., product)
probability measure on $\CantorSet$, every set that is
$\nu$-measurable in ${\cal C}$ and (like most complexity classes)
invariant under finite alterations must have $\nu$-measure $0$ or
$\nu$-measure $1$ in ${\cal C}$.

The theory is presented here is based on resource-bounded {\it
martingale splitting operators}, which are type-2 functionals, each of
which maps $\N \times {\cal D}_\nu$ into ${\cal D}_\nu \times {\cal
D}_\nu$, where ${\cal D}_\nu$ is the set of all $\nu$-martingales.
This type-2 aspect of the theory appears to be essential for general
$\nu$-measure in complexity classes ${\cal C}$, but the sets of
$\nu$-measure $0$ or $1$ in ${\cal C}$ are shown to be characterized
by the success conditions for martingales (type-1 functions) that have
been used in resource-bounded measure to date.
\end{abstract}

\newpage
\pagenumbering{arabic}
\pagestyle{plain}
\section{Introduction}

Resource-bounded measure is a complexity-theoretic generalization of
classical measure theory that interacts informatively and
quantitatively  with many other much-studied aspects of computational
complexity.   Since its introduction in 1992 \cite{Lutz:AEHNC},
resource-bounded measure has yielded a rapidly growing body of new
results and insights across a variety of subareas of computational
complexity.  The recent survey papers \cite{Lutz:QSET,AmbMay97} summarize
many of these developments, but the ongoing progress of many
investigators is quickly outdating these surveys.

Notwithstanding its productivity to date, the theory developed in
\cite{Lutz:AEHNC} can only be regarded as a fragment of resource-bounded
measure.  There are two compelling reasons for this view.  First, the
theory in \cite{Lutz:AEHNC} is restricted to the case where the underlying
probability measure is the {\it uniform} probability measure on the
Cantor space $\CantorSet$ of all languages, i.e., the random experiment
in which a language $A \subseteq \strings$ is chosen by using an {\it
independent} toss of a {\it fair} coin to decide membership of each
string in $A$.  Second, the theory in \cite{Lutz:AEHNC} is restricted to
the sets of resource-bounded measure $0$ (and their complements, the
sets of resource-bounded measure $1$).

Recent work of Breutzmann and Lutz \cite{Lutz:EMCC}, furthered by Kautz
\cite{Kaut97}, has addressed the first of these restrictions by
investigating the measure-zero fragment of resource-bounded
$\nu$-measure, where the underlying probability measure $\nu$ is an
arbitrary (Borel) probability measure on $\CantorSet$.  Specific
results may require $\nu$ to satisfy certain complexity, independence,
or positivity conditions, but these requirements arise naturally, not
as artifacts of the theory.

In this paper, we address the second of the above-mentioned
restrictions by developing a general theory of resource-bounded
measurability and measure.  Starting from any feasible probability
measure $\nu$ on $\CantorSet$ and any suitable class ${\cal C} \subseteq
\CantorSet$, we identify the subsets of $\CantorSet$ that are
$\nu$-measurable in ${\cal C}$ and assign $\nu$-measures to these
sets, thereby endowing ${\cal C}$ with internal measure-theoretic
structure.  Classes ${\cal C}$ to which the theory applies include
various exponential time and space complexity classes, the class of
all decidable languages, and the Cantor space $\CantorSet$ itself.  We
show that classical $\nu$-measure on $\CantorSet$ is precisely the case
${\cal C} = \CantorSet$ of the more general theory presented here.

We show that the sets that are $\nu$-measurable in ${\cal C}$ form an
algebra relative to which $\nu$-measure in ${\cal C}$ is well-behaved
(monotone, additive, etc.).  We also show that this algebra is
complete (subsets of measure $0$ sets are measurable), that it is
closed under sufficiently uniform infinitary unions and intersections,
and that $\nu$-measure in ${\cal C}$ has the appropriate additivity
and monotone convergence properties with respect to these infinitary
operations.

We prove a resource-bounded generalization of the classical Kolmogorov
zero-one law \cite{Kolm33}.  This states that, when $\nu$ is a
feasible coin-toss (i.e., product) probability measure on $\CantorSet$,
every set that is $\nu$-measurable in ${\cal C}$ and is (like most
complexity classes) invariant under finite alterations must have
$\nu$-measure $0$ in ${\cal C}$ or $\nu$-measure $1$ in ${\cal C}$.

The theory presented here is not a straightforward extension of the
measure-zero fragment of resource-bounded measure that has been
investigated to date.  The basic objects in our development here are
resource-bounded {\it martingale splitting operators}, each of which
is a type-2 functional mapping $\N \times {\cal D}_\nu$ into ${\cal
D}_\nu \times {\cal D}_\nu$, where ${\cal D}_\nu$ is the set of all
$\nu$-martingales.  This approach, which can be regarded as a
martingale-based, complexity-theoretic generalization of the classical
Carath\'{e}odory definition of measure, entails a variety of new proof
techniques.  This type-2 aspect of the theory appears to be essential
in order to achieve the algebraic properties mentioned above in
complexity classes.  However, we show that the sets of $\nu$-measure
$0$ or $1$ in ${\cal C}$ are characterized by the same success
conditions for martingales (type-1 functions) that have been used in
resource-bounded measure to date.

The theory presented here extends and deepens the relationship between
resource-bounded measure and classical measure theory.  We expect this
to open the way for new applications in computational complexity by
enabling the adaptation of powerful techniques from measure-theoretic
probability theory.  The theory also opens the way for the
investigation of complexity classes using probability measures that
are not subject to the Kolmogorov zero-one law, thereby accruing the
quantitative benefit of measures throughout the interval $[0, 1]$.
This may lead to new applications of the resource-bounded
probabilistic method, especially in situations where the application
itself may dictate the use of such a probability measure.  Finally, we
believe that the work presented here highlights the importance of the
continuing investigation of higher-type computational complexity.

\section{Notation and Functionals}
\label{Prelim}

In this paper, $\N$ is the set of nonnegative integers, $\Q$ is
the set of rational numbers, and $\R$ is the set of real
numbers.

We write $\strings$ for the set of all (finite, binary) {\it strings},
and we write $|x|$ for the length of a string $x$.  The empty string,
$\lambda$, is the unique string of length 0.  The {\it standard
  enumeration} of $\strings$ is the sequence $s_0 = \lambda, s_1 = 0, s_2
= 1, s_3 = 00, \ldots$, ordered first by length and then
lexicographically.  For $x, y \in \strings$, we write $x < y$ if $x$
precedes $y$ in this standard enumeration.  For $n \in \N$, $\{0,
1\}^n$ denotes the set of all strings of length $n$, and $\{0,
1\}^{\leq n}$ denotes the set of all strings of length at most $n$.

If $x$ is a string or an (infinite, binary) {\it sequence}, and if $0
\leq i \leq j < |x|$, then $x[i..j]$ is the string consisting of the
$i^{\rm th}$ through $j^{\rm th}$ bits of $x$.  In particular, $x[0..i-1]$
 is the $i$-{\it bit prefix} of $x$.  We write
$x[i]$ for $x[i..i]$, the $i^{\rm th}$ bit of $x$.  (Note that
the leftmost bit of $x$ is $x[0]$, the $0^{\rm th}$ bit of $x$.)

If $w$ is a string and $x$ is a string or sequence, then we write $w
\sqsubseteq x$ if $w$ is a prefix of $x$, i.e., if there is a string
or sequence $y$ such that $x = wy$.

The {\it Boolean value} of a condition $\phi$ is $\bval{\phi} =$ {\bf
  if} $\phi$ {\bf then} 1 {\bf else} 0.

We use the discrete logarithm
$$\log n = \min \{ k\in \N | 2^k \geq n \}.$$
Note that $\log 0 = 0$.

As in \cite{Lutz:AEHNC}, for each $i \in \N$ we define a class $G_i$  of functions from $\N$ into $\N$ as follows.
\begin{align*}
G_0& = \{ f | (\exists k) f(n) \leq kn \ \} \\
G_{i+1} &= 2^{G_i (\log n)} = \{ f | (\exists g \in
   G_i) f(n) \leq 2 ^{g(\log n)}\ \}
\end{align*}
We also define the functions $\hat{g}_i \in G_i$ by
$\begin{array}{ccc}
\hat{g}_0(n) = 2n, & & \hat{g}_{i+1}(n) = 2^{\hat{g}_i(\log n)}.
\end{array}$
We regard the functions in these classes as growth rates.  In particular,
$G_0$ contains the linearly bounded growth rates and $G_1$ contains the
polynomially bounded growth rates. It is easy to show that each $G_i$ is closed
under composition, that each $f\in G_i$ is $o(\hat{g}_{i+1})$, and that
each $\hat{g}_i$ is $o(2^n)$.  Thus $G_i$ contains superpolynomial growth rates
for all $i>1$, but all growth rates in the $G_i$-hierarchy are subexponential.

Within the class REC of all decidable languages, we are interested in the
uniform complexity classes E$_i = $ DTIME($2^{G_{i-1}}$) and
E$_i$SPACE = DSPACE ($2^{G_{i-1}}$) for $i \geq 1$.  The
well-known exponential complexity classes
E = E$_1$ = DTIME(2$^{\mbox{linear}}$),
E$_2$ = DTIME(2$^{\mbox{polynomial}}$), ESPACE = E$_1$SPACE = DSPACE
(2$^{\mbox{linear}}$), and E$_2$SPACE = DSPACE(2$^{\mbox{polynomial}}$) are of
particular interest.

In this extended abstract, our discussion of functions, functionals,
and their complexities is somewhat informal and is presented in terms
of examples.

Many of the functions in this paper are real-valued functions on
discrete domains.  A typical example might have the form
\[
\func{f}{\N \times \strings}{\R},
\]
and we often write $f_k(w)$ for $f(k, w)$.  A {\it computation} of
such a function $f$ is a function
\[
\func{\widehat{f}}{\N \times \N \times \strings}{\Q}
\]
such that, for all $r, k \in \N$ and $w \in \strings$,
\[
\left| \widehat{f}_{r, k}(w) - f_k(w) \right| < 2^{-r}.
\]
The {\it canonical computation} of such a function $f$ is the unique
computation $\widehat{f}$ of $f$ such that, for all $r, k \in \N$ and
$w \in \strings$, $\widehat{f}_{r, k}(w)$ is of the form $a \cdot
2^{-k}$, where $a$ is an integer.

Given a set $\Delta$, we say that $f$ is $\Delta$-{\it{computable}} if
there exists $\widehat{f} \in \Delta$ such that $\widehat{f}$ is a
computation of $f$.  For the sets $\Delta$ that we consider in this
paper, this is equivalent to saying that the canonical computation of
$f$ is an element of $\Delta$.

We also consider {\it functionals}, whose arguments and values may
themselves be real-valued functions.  A typical example might have the
form
\[
\func{\Phi}{\N \times (\strings \longrightarrow \R)}{(\strings
\longrightarrow \R)},
\]
in which case for each $k \in \N$ and $\func{f}{\strings}{\R}$, the
value of $\Phi$ is a function $\func{\Phi_k(f)}{\strings}{\R}$.
Formally, we regard such a functional as operating not on the
real-valued functions themselves, but rather on {\it canonical
computations} of
these functions, as defined in the preceding paragraph.  Thus we
identify the example functional $\Phi$ above with (any) functional
\[
\func{\Phi^\prime}{\N \times (\N \times \strings \longrightarrow
\Q)}{(\N \times \strings \longrightarrow \Q)}
\]
such that, for all $k \in \N$ and $\func{f}{\strings}{\R}$, if
$\widehat{f}$ is the canonical computation of $f$, then
$\Phi^\prime_k(\widehat{f})$ is
a computation of $\Phi_k(f)$.  Note that $\Phi^\prime$ is a type-2
functional; it is in this sense that we regard $\Phi$ as a type-2
functional.

If we wish to discuss the computability or complexity of the
functional $\Phi$ above, then we curry the functional $\Phi^\prime$,
obtaining a functional
\[
\func{\Phi^{\prime \prime}}{\N \times (\N \times \strings
\longrightarrow \Q) \times \N \times \strings}{\Q}
\]
We then say that $\Phi^\prime$ is {\it computable} if there is a
function oracle Turing machine $M$ such that, for all $k \in \N$,
$\func{g}{\N \times \strings}{\Q}$, $r \in \N$, and $w \in \strings$,
$M$ on input $(k, r, w)$ with oracle $g$ computes the value
$\Phi^{\prime \prime}(k, g, r, w) = \Phi^\prime_k(g)_r(w)$  We say
that $\Phi$ is {\it computable} if $\Phi$ is, in the above manner,
identified with some computable functional $\Phi^\prime$.  Similarly,
we say that $\Phi$ is {\it computable in time} $t(n)$ if there is an
oracle Turing machine $M$ as above that runs in at most $t(k+r+|w|)$
steps, and we say that $\Phi$ is {\it computable in space} $s(n)$ if
there is an oracle Turing machine $M$ as above that uses at most
$s(k+r+|w|)$ tape cells, {\it including} cells used for oracle queries
and output.

We use the following classes of functionals.
\begin{enumerate}
\item[1.] The class ``all,'' consisting of all functionals of type
$\leq 2$ in the above sense.
\item[2.] The class
\[
{\rm \rec} = \set{\Phi \in {\rm all}}{\Phi \text{ is computable}}.
\]
\item[3.] For each $i \geq 1$, the class
\[
{\rm p}_i = \set{ \Phi \in {\rm all}}{\Phi \text{ is computable in
$G_i$ time}}.
\]
\item[4.] For each $i \geq 1$, the class
\[
{\rm p}_i{\rm space} = \set{ \Phi \in {\rm all}}{\Phi \text{ is
computable in $G_i$ space}}.
\]
\end{enumerate}
We write ${\rm p}$ for ${\rm p}_1$ and pspace for ${\rm
p}_1{\rm{space}}$.  Throughout this paper, a {\it resource bound}
(generically denoted by $\Delta$ or $\Delta^\prime$) is one of the
above classes of functionals.

As in \cite{Lutz:AEHNC}, a {\it constructor} is a function
$\func{\delta}{\strings}{\strings}$ such that $x \psqsubset \delta(x)$
for all $x \in \strings$.  The {\it result} of a constructor $\delta$
is the unique language $R(\delta)$ such that $\delta^k(\lambda)
\sqsubseteq R(\delta)$ for all $k \in \N$, where $\delta^k$ is the
$k$-fold composition of $\delta$ with itself.  The {\it result class}
of a resource bound $\Delta$ is the set of all languages $R(\delta)$
such that $\delta$ is a constructor and $\delta \in \Delta$.  As noted
in \cite{Lutz:AEHNC}, we have
\begin{align*}
&R({\rm all}) = \CantorSet, \\
&R({\rm rec}) = \REC, \\
&R({\rm p}_i) = {\rm E}_i, \\
&R({\rm p}_i({\rm{space}}) = {\rm E}_i{\rm{SPACE}}
\end{align*}
for all $i \geq 1$.

\section{Martingales}
\label{Martingales}
We work in the {\it Cantor space} $\CantorSet$, consisting of all
languages (i.e., decision problems) $A \subseteq \strings$.  We
identify each language $A$ with its {\it characteristic sequence}, which is
the infinite binary sequence $A$ whose $n^{\rm th}$ bit is $A[n] =
\bval{s_n \in A}$ for each $n \in \N$.  Relying on this
identification, we also consider $\CantorSet$ to be the set of all
infinite binary sequences.

For each string $w \in \strings$, the {\it cylinder generated by} $w$
is the set
\[
\CantorSet_w = \set{A \in \CantorSet}{w \sqsubseteq A}
\]
Note that $\CantorSet_{\lambda} = \CantorSet$.

A {\it probability measure} on $\CantorSet$ is a function
\[
\func{\nu}{\strings}{[0, 1]}
\]
such that $\nu(\lambda) = 1$ and, for all $w \in \strings$,
\[
\nu(w) = \nu(w0) + \nu(w1).
\]
Intuitively, $\nu(w)$ is the probability that $A \in \CantorSet_w$
when we ``choose a language $A \in \CantorSet$ according to the
probability measure $\nu$.''  We sometimes write $\nu(\CantorSet_w)$
for $\nu(w)$.

If $v, w \in \strings$ and $\nu(w) > 0$, then we write
\[
\nu(v \mid w) =
\begin{cases}
1 & \text{if $v \sqsubseteq w$} \\
& \\
\frac{\nu(v)}{\nu(w)} & \text{if $w \sqsubseteq v$} \\
& \\
0 & \text{otherwise}
\end{cases}
\]
for the {\it conditional} $\nu$-{\it{measure}} of $v$ given $w$.

The {\it uniform probability measure} $\mu$ is defined by
\[
\mu(w) = 2^{-|w|}
\]
for all $w \in \strings$.

A {\it bias sequence} is a sequence $\bvec = (\beta_k \mid k \in \N)$,
where each $\beta_k \in [0, 1]$.  Given a bias sequence $\bvec$, the
$\bvec$-{\it{coin-toss probability measure}} (also called the
$\bvec$-{\it{product probability measure}}) is the probability measure
$\mu^{\bvec}$ defined by
\[
\mu^{\bvec}(w) = \prod_{k = 0}^{|w| - 1} \left[(1 - \beta_k) \cdot (1 - w[k])
+ \beta_k \cdot w[k]\right]
\]
for all $w \in \strings$.  Intuitively, $\mu^\bvec(w)$ is the
probability that $w \sqsubseteq A$ when the language $A \subseteq
\strings$ is chosen according to the following random experiment.  For
each string $s_k$ in the standard enumeration $(s_k \mid k \in \N)$ of
$\strings$, we (independently of all other strings) toss a special
coin, whose probability is $\beta_k$ of coming up heads, in which case
$s_k \in A$, and $1 - \beta_k$ of coming up tails, in which case $s_k
\not \in A$.  Note that, in the special case where $\beta_k =
\frac{1}{2}$ for all $k \in \N$, $\mu^{\bvec}$ is the uniform
probability measure $\mu$.

If $\Delta$ is a resource bound, as specified in section~\ref{Prelim},
then a $\Delta$-{\it{probability measure}} on $\CantorSet$ is a
probability measure $\nu$ on $\CantorSet$ with the following two
properties.
\begin{enumerate}
\item[(i)] $\nu$ is $\Delta$-computable.
\item[(ii)] There is a $\Delta$-computable function $\func{l}{\N}{\N}$
such that, for all $w \in \strings$, $\nu(w) = 0$ or $\nu(w) \geq
2^{-l(|w|)}$.
\end{enumerate}
Note that, if $\nu$ is a $\Delta$-probability measure on $\CantorSet$,
then the Boolean value $\bval{ \nu(w) = 0 }$ is $\Delta$-computable.

We now recall the well-known notion of a martingale over a probability
measure $\nu$.  Computable martingales were used by Schnorr
\cite{Schn70,Schn71a,Schn71b,Schn73} in his investigations of randomness, and have more
recently been used by Lutz \cite{Lutz:AEHNC} and Breutzmann and Lutz
\cite{Lutz:EMCC} in the development of resource-bounded measure.

If $\nu$ is a probability measure on $\CantorSet$, then a
$\nu$-{\it{martingale}} is a function $\func{d}{\strings}{[0,
\infty)}$ such that, for all $w \in \strings$,
\[
d(w)\nu(w) = d(w0)\nu(w0)+d(w1)\nu(w1).
\]
The {\it initial value} of a $\nu$-martingale $d$ is $d(\lambda)$.  A
$\Delta$-$\nu$-{\it{martingale}} is a $\nu$-martingale that is
$\Delta$-computable.  We reserve the symbol $\one$ for the {\it unit
martingale} define by $\one(w) = w$ for all $w \in \strings$.  Note
that $\one$ is a $\nu$-{martingale} for every probability measure
$\nu$ on $\CantorSet$.

Let $d$ be a $\nu$-martingale, and let $A \subseteq \strings$.  We say
that $d$ {\it covers} $A$ if there exists $n \in \N$ such that
$d(A[0..n-1]) \geq 1$.  We say that $d$ {\it succeeds on} $A$ if
\[
\limsup_{n \rightarrow \infty} d(A[0..n-1]) = \infty.
\]
We say that $d$ {\it succeeds strongly on} $A$ if
\[
\lim_{n \rightarrow \infty} d(A[0..n-1]) = \infty.
\]
The {\it set covered by} $d$ is the set
\[
S^1[d] = \set{A}{d \text{ covers } A}.
\]
The {\it success set} of $d$ is
\[
S^\infty[d] = \set{A}{d \text{ succeeds on } A}.
\]
The {\it strong success set} of $d$ is
\[
S^\infty_{\rm str}[d] = \set{A}{d \text{ succeeds strongly on } A}.
\]
The set $S^1[d]$ is also called the {\it unitary success set} of $d$.

A {\it prefix set} is a language $A \subseteq \strings$ with the
property that no element of $A$ is a prefix of another element of
$A$.  A routine induction on the definition of martingales yields the
following.

\begin{lemma}
\label{G.1}
If $d$ is a $\nu$-martingale, then for every prefix set $A \subseteq
\strings$,
\[
\sum_{w \in A} d(w)\nu(w) \leq d(\lambda).
\]
\end{lemma}

An {\it open set} in $\CantorSet$ is any set of the form
\[
\CantorSet_A = \bigcup_{w \in A} \CantorSet_w
\]
for $A \subseteq \strings$.  Every open set can in fact be written in
the form $\CantorSet_A$, where $A$ is a prefix set.  This
representation is not unique, but if $A$ and $B$ are prefix sets such
that $\CantorSet_A = \CantorSet_B$, then it is easy to see that
\[
\sum_{w \in A} \nu(w) = \sum_{w \in B} \nu(w)
\]
for every probability measure $\nu$ on $\CantorSet$.  The
$\nu$-{\it{measure}} of an open set $X \subseteq \CantorSet$, given by
\[
\nu(X) = \sum_{w \in A} \nu(w)
\]
where $A$ is any prefix set such that $X = \CantorSet_A$, is thus
well-defined.

If $d$ is any $\nu$-martingale, then the set $S^1[d]$ is open because
$S^1[d] = \CantorSet_A$, where $A = \set{w}{d(w) \geq 1}$.  Thus the
$\nu$-measure $\nu(S^1[d])$ is a well-defined real number.  Before
proceeding, however, we note that $\nu(S^1[d])$ may fail to be
computable, even when $d$ is ${\rm p}$-computable.

\begin{example}
\label{G.2}
Let $(M_i \mid i \in \N)$ be a standard enumeration of Turing
machines, and let
\[
K = \set{i \in \N}{M_i(0^i) \text{ halts}}
\]
be the diagonal halting problem.  For each $i \in K$, let $t(i)$ be
the number of steps executed by $M_i(0^i)$, and let $T_i =
\set{0^i1u}{u \in \{0, 1\}^{t(i)}}$.  Let $T = \bigcup_{i \in K} T_i$,
and note that $T$ is a prefix set.  Define
$\func{d}{\strings}{[0, \infty)}$ by
\[
d(w) =
\begin{cases}
1 & \text{if $v1 \sqsubseteq w$ for some $v \in T$} \\
& \\
0 & \text{if $v0 \sqsubseteq w$ for some $v \in T$} \\
& \\
\frac{1}{2} & \text{otherwise}.
\end{cases}
\]
It is easy to check that $d$ is a ${\rm p}$-$\mu$-martingale, where
$\mu$ is the uniform probability measure on $\CantorSet$.  It is also
easy to see that
\[
\mu(S^1[d]) = \frac{1}{4} \sum_{i \in K} 2^{-i}.
\]
It follows that $\mu(S^1[d])$ is Turing-equivalent to $K$, and hence
not computable.  (In fact, $\mu(S^1[d])$ is a version of Chaitin's
random real number $\Omega$ \cite{Chai75a}.)
\end{example}

Even though $\nu(S^1[d])$ may not be computable, it is always bounded
above by the initial value $d(\lambda)$.

\begin{lemma}
\label{G.3}
For every $\nu$-martingale $d$,
\[
\nu(S^1[d]) \leq d(\lambda).
\]
\end{lemma}
\begin{proofof}[Lemma~\ref{G.3}]
Fix a prefix set $A \subseteq \set{w}{d(w) \geq 1}$ such that $S^1[d]
= \CantorSet_A$.  Then
\[
\nu(S^1[d]) = \sum_{w \in A} d(w) \nu(w) \leq d(\lambda)
\]
by Lemma~\ref{G.1}
\end{proofof}

If $d$ is $\Delta$-computable and $X \subseteq S^1[d]$, then
Lemma~\ref{G.3} says that we can regard $d$ as an explicit,
$\Delta$-computable certification that $X$ does not have $\nu$-measure
greater than the initial value $d(\lambda)$, which is itself
$\Delta$-computable.

By Lemma~\ref{G.3}, no $\nu$-martingale $d$ can cover a cylinder
$\CantorSet_w$ unless $d(\lambda)$ is at least $\nu(w)$.  The
following theorem, which is central to resource-bounded measure, says
that, for $\Delta$-$\nu$-martingales $d$, this remains true even if we
intersect the cylinder with $R(\Delta)$.

\begin{theorem}[Measure Conservation Theorem]
\label{G.4}
If $w \in \strings$ and $d$ is a $\Delta$-$\nu$-martingale such that
$\CantorSet_w \cap R(\Delta) \subseteq S^1[d]$, then $d(\lambda) \geq
\nu(w)$.
\end{theorem}
\begin{proofof}[Theorem~\ref{G.4}]
Assume that $w \in \strings$ and $d$ is a $\Delta$-$\nu$-martingale
such that $d(\lambda) < \nu(w)$.  It suffices to exhibit a constructor
$\delta \in \Delta$ such that $R(\delta) \in \CantorSet_w - S^1[d]$.

First note, for every prefix $w^\prime \sqsubseteq w$, the definition
of $\nu$-martingales(applied inductively) tells us that
\begin{align*}
d(w^\prime)\nu(w^\prime) &\leq \sum_{|u| = |w^\prime|} d(u)\nu(u) =
d(\lambda)\nu(\lambda) \\
&= d(\lambda) < \nu(w) \leq \nu(w^\prime),
\end{align*}
so $d(w^\prime) < 1$.  In particular, then fix a constant $m \in \N$ such that $d(w) \leq 1 - 2^{1-m}$.

Let $\widehat{d}$ be a $\Delta$-computation
of $d$.  Using $\widehat{d}$ and the constants $w$ and $m$, define
$\func{\delta}{\strings}{\strings}$ by
\[
\delta(x) =
\begin{cases}
w & \text{if $x \psqsubset w$} \\
& \\
x0 & \text{if $\widehat{d}_{a(x)}(x0) \leq \widehat{d}_{a(x)}(x1)$ and
note $x \psqsubset w$} \\
& \\
x1 & \text{otherwise},
\end{cases}
\]
where $a(x) = |x| + m + 2$.  It is clear that $\delta$ is a
constructor, $\delta \in \Delta$, and $R(\delta) \in \CantorSet_w$.

For any string $x$ such that $w \sqsubseteq x$, the definition of
$\delta$ and the fact that $d$ is a $\nu$-martingale ensure that
\begin{align*}
d(\delta(x)) &\leq \widehat{d}_{a(x)}(\delta(x)) + 2^{-a(x)} \\
&= \min_{b \in \{0, 1\}} \widehat{d}_{a(x)}(xb) + 2^{-a(x)} \\
&\leq \min_{b \in \{0, 1\}} d(xb) + 2^{1 - a(x)} \\
&\leq d(x) + 2^{1-a(x)} \\
&= d(x) + 2^{-(|x| + m + 1)}.
\end{align*}
It follows that, for all $k \in \N$,
\begin{align*}
d(\delta^k(w)) &\leq d(w) + \sum_{j=0}^{k-1} 2^{-(j + m + 1)} \\
&< d(w) + 2^{-m} \\
&\leq 1 - 2^{-m}.
\end{align*}
Since every prefix of $R(\delta)$ is either a prefix of $w$ or of the
form $\delta^k(w)$ for some $k \in \N$, we have now shown that $R(\delta)
\not \in S^1[d]$, completing the proof.
\end{proofof}

A $\nu$-martingale is {\it regular} if, for all $v, w \in \strings$, if
$\nu(v) \geq 1$ and $v \sqsubseteq w$, then $\nu(w) \geq 1$.  It is
often technically convenient to have a uniform means of ensuring that
martingales are regular.  The following lemma provides such a
mechanism.  Let $\Delta$ be a resource bound, as specified in
section~\ref{Prelim}, and let $\nu$ be a probability measure on
$\CantorSet$.

\begin{lemma}[Regularity Lemma]
\label{G.5}
There is a functional
\[
\func{\Lambda}{{\cal D}_\nu}{{\cal D}_{\nu}}
\]
with the following properties.
\begin{enumerate}
\item[1.] For all $d \in {\cal D}_\nu$, $\Lambda(d)$ is a regular
$\nu$-martingale such that $\Lambda(d)(\lambda) = d(\lambda)$ and
$S^1[d] \subseteq S^1[\Lambda(d)]$.
\item[2.] $\Lambda(\one) = \one$.
\item[3.] If $\nu$ is a $\Delta$-probability measure on $\CantorSet$,
then $\Lambda$ is $\Delta$-computable.
\end{enumerate}
\end{lemma}
\begin{proofof}[Lemma~\ref{G.5}](sketch)
Given $\alpha \in (0, 1)$, it is convenient to have notations for the $\alpha$-weighted averaging function
\[
\func{m_\alpha}{\R^2}{\R}
\]
\[
m_\alpha(s, t) = \alpha s + (1 -\alpha)t,
\]
the half-plane
\[
H_\alpha = \set{(s, t) \in \R^2}{m_\alpha(s, t) \geq 1}
\]
and the region
\[
D_\alpha = H_\alpha \cup [0, \infty)^2.
\]
The ``Robin Hood function''
\[
\func{rh_\alpha}{D_\alpha}{[0, \infty)^2}
\]
is then defined as follows.
\begin{enumerate}
\item[(i)] If $(s, t) \in [0, 1]^2$, then
\[
rh_\alpha(s, t) = (s, t).
\]
\item[(ii)] If $(s, t) \in H_\alpha$, then
\[
rh_\alpha(s, t) = (m_\alpha(s, t), m_\alpha(s, t)).
\]
\item[(iii)] If $(s, t) \in D_\alpha - H_\alpha$ and $s \geq 1$, then
\[
rh_\alpha(s, t) = (1, \frac{m_\alpha(s-1, t)}{1 - \alpha}).
\]
\item[(iv)] If $(s, t) \in D_\alpha - H_\alpha$ and $t \geq 1$, then
\[
rh_\alpha(s, t) = (\frac{m_\alpha(s, t-1)}{\alpha}, 1).
\]
\end{enumerate}
We also use the notation $rh_\alpha(s, t) = (rh^{(0)}_\alpha(s, t), rh^{(1)}_\alpha(s, t))$.

The following essential properties of the Robin Hood function $rh_\alpha$ are routine to verify.

\begin{enumerate}
\item[1.] The transformation $rh_\alpha$ is a continuous, piecewise linear mapping from $D_\alpha$ into $[0, \infty)^2$.
\item[2.] The transformation $rh_\alpha$ preserves $\alpha$-weighted averages, i.e., $m_\alpha(rh_\alpha(s, t)) = m_\alpha(s, t)$ for all $(s, t) \in D_\alpha$.
\item[3.] The transformation $rh_\alpha$ maps $H_\alpha$ into $[1,
\infty)^2$.  That is, if the average $m_\alpha(s, t)$ is at least $1$,
then $rh_\alpha$ ``steals from the richer and gives to the poorer'' of
$s$ and $t$ so that both $rh^{(0)}_\alpha(s, t)$ and
$rh^{(1)}_\alpha(s, t)$ are at least $1$.
\item[4.] For all $(s, t) \in D_\alpha$, $rh^{(0)}_\alpha(s, t) \geq
\min \{1, s \}$ and $rh^{(1)}_\alpha(s, t) \geq \min \{1, t \}$.  That
is, the transformation $rh_\alpha$ never ``steals'' more than the
excess above $1$.
\item[5.] The transformation $rh_\alpha$ leaves points of $[0, 1]^2$ unchanged.
\end{enumerate}

Using the Robin Hood function, we define the functional
$\func{\Lambda}{{\cal D}_\nu}{{\cal D}_\nu}$ as follows.  For $d \in
{\cal D}_\nu$, we define the $\nu$-martingale $\Lambda(d)$ by the
following recursion.  (In all clauses, $w \in \strings$ and $b \in
\{0, 1\}.$)
\begin{enumerate}
\item[(i)] $\Lambda(d)(\lambda) = d(\lambda)$.
\item[(ii)] If $\nu(w) = 0$ or $\nu(wb \mid w) \in \{0, 1\}$, then
$\Lambda(d)(wb) = \Lambda(d)(w)$.
\item[(iii)] If $\nu(w) > 0$ and $0 < \nu(wb \mid w) < 1$, then
\[
\Lambda(d)(wb) = rh^{(b)}_{\nu(w0 \mid w)} (g_0(w), g_1(w)),
\]
where $g_b(w) = \Lambda(d)(w) - d(w) + d(wb)$.
\end{enumerate}
It is now routine (if tedious) to verify that $\Lambda$ has the desired properties.
\end{proofof}

\section{Measure}
\label{Measure}

In this, the main section of the paper, we develop the general theory
of resource-bounded measurability and measure in complexity classes.
Throughout this paper, $\Delta$ and $\DeltaPrime$ are resource bounds,
as specified in section~\ref{Prelim}, and $\nu$ is a probability
measure on $\CantorSet$, as defined in section~\ref{Martingales}.

The fundamental objects in resource-bounded measure are {\it
martingale splitting operators}.  If $\Xplus$ and $\Xminus$ are
disjoint subsets of $\CantorSet$, then a martingale splitting operator
for $\Xplus$ and $\Xminus$ is a type-2 functional that, given a
martingale $d$ and error tolerance $2^{-r}$, ``splits'' $d$ into two
martingales $\dpr$ and $\dmr$ such that $\dpr$ covers every element of
$\Xplus$ covered by $d$, $\dmr$ covers every element of $\Xminus$
covered by $d$, and the sum of initial values $\dpr(\lambda) +
\dmr(\lambda)$ does not exceed the initial value $d(\lambda)$ by
more than $2^{-r}$.  The formal definition follows.

\begin{definition*}
If $\Xplus$ and $\Xminus$ are disjoint subsets of $\CantorSet$, then a
$\nu$-{\it splitting operator} for $(\Xplus, \Xminus)$ is a functional
\[
\func{\Phi}{\N \times \distnu}{\distnu \times \distnu},
\]
where we write
\[
\Phi(r, d) = (\Phi^+_r(d), \Phi^-_r(d)),
\]
such that the following conditions hold for all $r \in \N$ and $d \in
\distnu$.
\begin{enumerate}
\item[(i)] $\Xplus \cap S^1[d] \subseteq S^1[\Phi^+_r(d)]$,
\item[(ii)] $\Xminus \cap S^1[d] \subseteq S^1[\Phi^-_r(d)]$,
\item[(iii)] $\Phi^+_r(d)(\lambda) + \Phi^-_r(d)(\lambda) \leq
d(\lambda) + 2^{-r}$.
\end{enumerate}
A $\Delta$-$\nu$-{\it{splitting operator}} for $(\Xplus, \Xminus)$ is
a $\nu$-splitting operator for $(\Xplus, \Xminus)$ that is
$\Delta$-computable.
\end{definition*}

The resource-bounded measurement of sets, both globally and in
$R(\Delta)$, is now defined in terms of splitting operators.

\begin{definition*}
Let $X \subseteq \CantorSet$.
\begin{enumerate}
\item[1.] A $\Delta$-$\nu$-{\it{measurement of}} $X$ is a
$\Delta$-$\nu$-splitting operator for $(X, X^{\rm c})$.
\item[2.] A $\nu$-{\it{measurement of}} $X$ {\it{in}} $R(\Delta)$ is a
$\Delta$-$\nu$-splitting operator for $(R(\Delta) \cap X, R(\Delta) -
X)$.
\end{enumerate}
\end{definition*}

If $\Yplus \subseteq \Xplus$ and $\Yminus \subseteq \Xminus$, then
every $\nu$-splitting operator for $(\Xplus, \Xminus)$ is clearly a
$\nu$-splitting operator for $(\Yplus, \Yminus)$.  In particular,
then, every $\Delta$-$\nu$-measurement of $X$ is a $\nu$-measurement
of $X$ in $R(\Delta)$.  As we shall see, the converse is (fortunately)
not true.

We now develop some basic properties of $\nu$-measurements.  Recall
the unit martingale $\one$ from section~\ref{Martingales}.

\begin{lemma}
\label{M.1}
Let $X \subseteq \CantorSet$.  If $\Phi$ and $\Psi$ are
$\nu$-measurements of $X$ in $R(\Delta)$, then for all $j, k \in \N$,
\[
\Phi^+_j(\one)(\lambda) + \Psi^-_k(\one)(\lambda) \geq 1.
\]
\end{lemma}
\begin{proofof}[Lemma~\ref{M.1}]
Assume the hypothesis, let $j, k \in \N$, and let
\[
d = \Phi^+_j(\one) + \Psi^-_k(\one).
\]
Then $d$ is a $\Delta$-$\nu$-martingale and
\begin{align*}
R(\Delta) &= (R(\Delta) \cap X) \cup (R(\Delta) - X) \\
& \subseteq S^1[\Phi^+_j(\one)] \cup S^1[\Psi^-_k(\one)] \\
& \subseteq S^1[d],
\end{align*}
so $d(\lambda) \geq 1$ by the Measure Conservation Theorem.
\end{proofof}

\begin{notation*}
If $\Phi$ is a $\nu$-splitting operator, then we write
\begin{align*}
\Phi^+_\infty &= \inf_{r \in \N} \Phi^+_r(\one)(\lambda), \\
\Phi^-_\infty &= \inf_{r \in \N} \Phi^-_r(\one)(\lambda).
\end{align*}
\end{notation*}

\begin{lemma}
\label{M.2}
Let $X \subseteq \CantorSet$.  If $\Phi$ is a $\nu$-measurement of $x$
in $R(\Delta)$, then $\Phi^+_\infty + \Phi^-_\infty = 1$ and, for all
$r \in \N$,
\[
\Phi^+_\infty \leq \Phi^+_r(\one)(\lambda) \leq \Phi^+_\infty + 2^{-r}
\]
and
\[
\Phi^-_\infty \leq \Phi^-_r(\one)(\lambda) \leq \Phi^-_\infty + 2^{-r}.
\]
Thus $\Phi^+_r(\one)(\lambda) \longrightarrow \Phi^+_\infty$ and
$\Phi^-_r(\one)(\lambda) \longrightarrow \Phi^-_\infty $ as $r
\longrightarrow \infty$.
\end{lemma}
\begin{proofof}[Lemma~\ref{M.2}]
Assume the hypothesis.  It follows immediately from
Lemma~\ref{M.1} that $\Phi^+_\infty + \Phi^-_\infty \geq 1$.
Also, for any $r \in \N$,
\begin{align*}
\Phi^+_\infty + \Phi^-_\infty &\leq \Phi^+_r(\one)(\lambda) +
\Phi^-_r(\one)(\lambda) \\
&\leq \one(\lambda) + 2^{-r} \\
&= 1 + 2^{-r},
\end{align*}
so $\Phi^+_\infty + \Phi^-_\infty \leq 1.$  Thus $\Phi^+_\infty +
\Phi^-_\infty = 1$.

Now fix $r \in \N$. Then, by the definitions of $\Phi^+_\infty$ and
$\Phi^-_\infty$, $\Phi^+_\infty \leq \Phi^+_r(\one)(\lambda)$ and
$\Phi^-_\infty \leq \Phi^-_r(\one)(\lambda)$. Also,
\begin{align*}
&(\Phi^+_r(\one)(\lambda) - \Phi^+_\infty) + (\Phi^-_r(\one)(\lambda)
- \Phi^-_\infty) \\
&= \Phi^+_r(\one)(\lambda) + \Phi^-_r(\one)(\lambda) - 1 \\
&\leq \one(\lambda) + 2^{-r} - 1 \\
&= 2^{-r},
\end{align*}
so $\Phi^+_r(\one)(\lambda) \leq \Phi^+_\infty + 2^{-r}$ and
$\Phi^-_r(\one)(\lambda) \leq \Phi^-_\infty + 2^{-r}$.
\end{proofof}

\begin{lemma}
\label{M.3}
Let $X \subseteq \CantorSet$.  If $\Phi$ and $\Psi$ are $\nu$-measurements
of $X$ in $R(\Delta)$, then for all $r \in \N$,
\[
\left| \Phi^+_r(\one)(\lambda) - \Psi^+_r(\one)(\lambda)
\right| \leq 2^{-r}
\]
and
\[
\left| \Phi^-_r(\one)(\lambda) - \Psi^-_r(\one)(\lambda)
\right| \leq 2^{-r}.
\]
Thus $\Phi^+_\infty = \Psi^+_\infty$ and $\Phi^-_\infty =
\Psi^-_\infty$.
\end{lemma}
\begin{proofof}[Lemma~\ref{M.3}]
Assume the hypothesis, and let $r \in \N$. Then, by Lemma
\ref{M.1},
\begin{align*}
1 &\leq \Phi^+_r(\one)(\lambda) + \Psi^-_r(\one)(\lambda) \\
  &\leq \Phi^+_r(\one)(\lambda) + (\one(\lambda) + 2^{-r} -
  \Psi^+_r(\one)(\lambda)),
\end{align*}
so $\Psi^+_r(\one)(\lambda) \leq \Phi^+_r(\one)(\lambda) + 2^{-r}$.
Similarly, $\Phi^+_r(\one)(\lambda) \leq \Psi^+_r(\one)(\lambda) +
2^{-r}$, so
\[
\left| \Phi^+_r(\one)(\lambda) - \Psi^+_r(\one)(\lambda)
\right| \leq 2^{-r}.
\]
The proof that
\[
\left| \Phi^-_r(\one)(\lambda) - \Psi^-_r(\one)(\lambda)
\right| \leq 2^{-r}
\]
is similar.
\end{proofof}

We now have the machinery we need to give unambiguous definitions of
resource bounded measure.

\begin{definition*}
A set $X \subseteq \CantorSet$ is $\nu$-{\it{measurable in}}
$R(\Delta)$, and we write $X \in {\cal F}^\nu_{R(\Delta)}$, if there
exists a $\nu$-measurement $\Phi$ of $X$ in $R(\Delta)$.  In this
case, the $\nu$-{\it{measure of }} $X$ {\it in} $R(\Delta)$ is the
real number $\nu(X \mid R(\Delta)) = \Phi^+_\infty$.  (By Lemma~\ref{M.3},
$\nu(X \mid R(\Delta))$ does not depend
on the choice of $\Phi$.)
\end{definition*}

\begin{definition*}
A set $X \subseteq \CantorSet$ is $\Delta$-$\nu$-{\it{measurable}},
and we write $X \in {\cal F}^\nu_\Delta$,if there exists a
$\Delta$-$\nu$-measurement $\Phi$ of $X$.  In this case, the
$\Delta$-$\nu$-{\it{measure}} of $X$ is the real number $\nu_\Delta(X)
= \Phi^+_\infty$.  (By Lemma~\ref{M.3},
$\nu_\Delta(X)$ does not
depend on the choice of $\Phi$.)
\end{definition*}

An intuitive remark on these definitions is in order here.  As noted
in section~\ref{Martingales}, a martingale $d$ that covers a set $X$ can be regarded
as an explicit certification that $X$ does not have measure greater
than the total value $d(\lambda)$.  Thus, if we apply a measurement
$\Phi$ of $X$ to the unit martingale $\one$, then the resulting
martingales $\Phi^+_r(\one)$ and $\Phi^-_r(\one)$, for $r \in \N$, can
be regarded collectively as an explicit certification that $X$ does
not have measure greater than $\Phi^+_\infty$ or less than $1 -
\Phi^-_\infty = \Phi^+_\infty$.  This is clearly a necessary condition
for measurability but we require further that the measurement $\Phi$
split arbitrary martingales (which may be more exotic and of
lesser total measure than $\one$) in analogous fashion.  The utility
of this requirement is evident in the proof of
Theorem~\ref{M.12}, where we establish that the measurable
sets form an algebra over which measure is additive.

In general, when we make an assertion of the form ``$\nu(X \mid
R(\Delta)) = \alpha$,'' it is implicit that $X$ is $\nu$-measurable in
$R(\Delta)$.  On the other hand, an assertion of the form ``$\nu(X
\mid R(\Delta) \neq \alpha$'' is the {\it negation} of ``$\nu(X \mid
R(\Delta)) = \alpha$,'' and thus means that {\it either} $X$ is not
$\nu$-measurable in $R(\Delta)$ {\it or} $\nu(X \mid R(\Delta)) =
\beta$ for some $\beta \neq \alpha$.  Similar remarks hold for the
assertions ``$\nu_{\Delta}(X) = \alpha$'' and ``$\nu_\Delta(X) \neq
\alpha$.''

The following two lemmas are now very obvious, but also very useful.

\begin{lemma}
\label{M.4}
Let $X \subseteq \CantorSet$.
\begin{enumerate}
\item[1.] If $X$ is $\nu$-computable in $R(\Delta)$, then for every
$\nu$-measurement $\Phi$ of $X$ in $R(\Delta)$ and every $r \in \N$,
\[
\nu(X \mid \R(\Delta)) \leq \Phi^+_r(\one)(\lambda) \leq \nu(X \mid
R(\Delta)) + 2^{-r}
\]
\item[2.] If $X$ is a $\Delta$-$\nu$-computable, then for every
$\Delta$-$\nu$-measurement $\Phi$ of $X$ and every $r \in \N$,
\[
\nu_\Delta(X) \leq \Phi^+_r(\one)(\lambda) \leq \nu_\Delta(X) +
2^{-r}.
\]
\end{enumerate}
\end{lemma}

\begin{lemma}
\label{M.5}
Let $X \subseteq \CantorSet$.
\begin{enumerate}
\item[1.] If $X$ is $\Delta$-$\nu$-measurable and $\Delta \subseteq
\DeltaPrime$, then $X$ is $\DeltaPrime$-$\nu$-measurable and
$\nu_{\DeltaPrime}(X) = \nu_\Delta(X)$.
\item[2.] If $X$ is $\Delta$-$\nu$-measurable, then $X$ is
$\nu$-measurable in $R(\Delta)$ and $\nu(X \mid R(\Delta)) = \nu_\Delta
(X)$.
\item[3.] $X$ is $\nu$-measurable in $R(\Delta)$ if and only if $X
\cap R(\Delta)$ is $\nu$-measurable in $R(\Delta)$, in which case
$\nu(X \mid R(\Delta)) = \nu( X \cap R(\Delta) \mid R(\Delta))$.
\end{enumerate}
\end{lemma}

Before proceeding further, we show that resource-bounded measure is a
generalization of classical measure theory.  More precisely, classical
measure theory is the special case $\Delta = {\rm all}$ of
resource-bounded measure.  We first recall the classical definitions
(specialized to the Cantor space $\CantorSet$.)

Recall the definitions of open sets and their measures
in section~\ref{Martingales}.

\begin{definition*}
The {\it outer} $\nu$-{\it{measure}} of a set $X \subseteq \CantorSet$
is
\[
\nu^\ast(X) = \inf \set{\nu(Y)}{Y \text{ is open and } X \subseteq Y}.
\]
\end{definition*}

Note that the outer $\nu$-measure $\nu^\ast(X)$ is defined for {\it
every} set $X \subseteq \CantorSet$, and that $0 \leq \nu^\ast(X) \leq
1$.  It is well known that outer measure is {\it subadditive}, i.e.,
that
\[
\nu^\ast(X \cup Y) \leq \nu^\ast(X) + \nu^\ast(Y)
\]
for all $X, Y \subseteq \CantorSet$.

\begin{definition*}
A set $X \subseteq \CantorSet$ is $\nu$-{\it{measurable}} if, for
every set $Y \subseteq \CantorSet$,
\[
\nu^\ast(Y) = \nu^\ast( Y \cap X) + \nu^\ast(Y - X).
\]
In this case, the $\nu$-{\it{measure}} of $X$ is
\[
\nu(X) = \nu^\ast(X).
\]
\end{definition*}

Since the outer measure is subadditive, it is evident that a set $X
\subseteq \CantorSet$ is $\nu$-measurable if, for every set $Y \subseteq
\CantorSet$,
\[
\nu^\ast(Y) \geq \nu^\ast(Y \cap X) + \nu^\ast( Y - X).
\]

We now show that classical measurability and measure are the special
case $\Delta = {\rm all}$ of resource-bounded measure.

\begin{theorem}
\label{M.6}
A set $X \subseteq \CantorSet$ is all-$\nu$-measurable if and only if
it is $\nu$-measurable, in which case $\nu_{\rm all}(X) = \nu(X)$.
\end{theorem}
\begin{proofof}[Theorem~\ref{M.6}]
Assume first that $X$ is all-$\nu$-measurable, and let $\Phi$ be an
all-$\nu$-measurement of $X$.  To see that $X$ is $\nu$-measurable,
let $Y \subseteq \CantorSet$ be arbitrary.  As noted above, to show
that $X$ is $\nu$-measurable, it suffices to prove that
\[
\nu^\ast(Y) \geq \nu^\ast(Y \cap X) + \nu^\ast(Y - X).
\]
For this purpose, let $r \in \N$ be arbitrary.  Then there is a prefix
set $A \subseteq \strings$ such that $Y \subseteq \CantorSet_A$ and
$\nu(\CantorSet_A) \leq \nu^\ast(Y) + 2^{-(r+1)}$.
Define $\func{d}{\strings}{[0, \infty)}$ by
\[
d(w) =
\begin{cases}
\sum_{u \in A} \nu(u \mid w ) & \text{if $\nu(w) > 0$} \\
& \\
1 & \text{if $\nu(w) = 0$}.
\end{cases}
\]
Then $d$ is a $\nu$-martingale and $Y \subseteq S^1[d]$, so
\[
Y \cap X \subseteq S^1[d] \cap X \subseteq S^1[\Phi^+_{r+1}(d)].
\]
Since $S^1[\Phi^+_{r+1}(d)]$ is open, it follows by Lemma
\ref{G.3} that
\[
\nu^\ast( Y \cap X) \leq \nu(S^1[\Phi^+_{r+1}(d)]) \leq
\Phi^+_{r+1}(d)(\lambda).
\]
A similar argument shows that
\[
\nu^\ast( Y - X ) \leq \Phi^-_{r+1}(d)(\lambda).
\]
We thus have
\begin{align*}
\nu^\ast( Y \cap X)+\nu^\ast(Y - X)
&\leq \Phi^+_{r+1}(d)(\lambda) + \Phi^-_{r+1}(d)(\lambda) \\
&\leq d(\lambda) + 2^{-(r+1)} \\
& = \nu(\CantorSet_A) + 2^{-(r+1)} \\
& \leq \nu^\ast(Y) + 2^{-r}.
\end{align*}
Since $r \in \N$ is arbitrary here, this shows that
\[
\nu^\ast(Y) \geq \nu^\ast(Y \cap X) + \nu^\ast( Y - X),
\]
confirming that $X$ is $\nu$-measurable.

If we let $Y = \CantorSet$ in the above argument, then $d = \one$ and
we have
\[
\nu(X) = \nu^\ast(\CantorSet \cap X) \leq \Phi^+_{r+1}(\one)(\lambda)
\]
for all $r \in \N$, so
\[
\nu(X) \leq \Phi^+_\infty = \nu_{\rm all}(X).
\]
Similarly,
\[
1 -\nu(X) = \nu(X^{\rm c}) = \nu^\ast( \CantorSet - X) \leq
\Phi^-_{r+1}(\one)(\lambda)
\]
for all $r \in \N$, so by Lemma~\ref{M.2},
\[
1 - \nu(X) \leq \Phi^-_\infty = 1 - \Phi^+_\infty = 1 - \nu_{\rm
all}(X),
\]
so $\nu_{\rm all}(X) = \nu(X)$.

Conversely, assume that $X$ is $\nu$-measurable.  Define
\[
\func{\Phi}{\N \times {\cal D}_\nu}{{\cal D}_\nu \times {\cal D}_\nu}
\]
as follows.  Let $r \in \N$ and $d \in {\cal D}_\nu$.  Since $S^1[d]$
is open, we know that $S^1[d]$, $S^1[d]
\cap X$ and $S^1[d] - X$ are $\nu$-measurable, with
\[
\nu(S[d]) = \nu(S[d] \cap X) + \nu(S[d] - X).
\]
Fix prefix sets $A^+$, $A^- \subseteq \strings$ such that
\begin{alignat*}{2}
S^1[d] \cap X & \subseteq \CantorSet_{A^+}, & \qquad
\nu(\CantorSet_{A^+}) \leq \nu(S[d] \cap X) &+ 2^{-(r+1)}, \\
S^1[d] - X & \subseteq \CantorSet_{A^-}, & \qquad
\nu(\CantorSet_{A^-}) \leq \nu(S[d] - X) & + 2^{-(r+1)}.
\end{alignat*}
For $a \in \{+, -\}$, define $\Phi^a_r(d)$ by
\[
\Phi^a_r(d)(w) =
\begin{cases}
\sum_{u \in A^a} \nu(u \mid w) & \text{if $\nu(w) > 0$} \\
& \\
1 & \text{if $\nu(w) = 0$.}
\end{cases}
\]
It is clear that the functional $\Phi$ so defined is all-computable
(a vacuous condition) and maps $\N \times {\cal D}_\nu$ into ${\cal
D}_\nu \times {\cal D}_\nu$.  To see that $\Phi$ is an
all-$\nu$-measurement of $X$, let
$r \in \N$ and $d \in {\cal D}_\nu$.  Three things are to be verified.
\begin{enumerate}
\item[1.] $S^1[d] \cap X \subseteq S^1[\Phi^+_r(d)]$.  To see this,
let $B \in S^1[d] \cap X$.  Then $B \in \CantorSet_{A^+}$, so there
exists $w \in A^+$ such that $w \sqsubseteq B$.  If $\nu(w) > 0$, then
\[
\Phi^+_r(d)(w) = \sum_{u \in A^+} \nu(u \mid w) \geq \nu( w \mid w) =
1,
\]
so $\Phi^+_r(d)(w) \geq 1$ in any case, whence $B \in
S^1[\Phi^+_r(d)]$.
\item[2.] $S^1[d] - X \subseteq S^1[\Phi^-_r(d)]$.  The verification
is analogous to that of 1.
\item[3.] $\Phi^+_r(d)(\lambda) + \Phi^-_r(d)(\lambda) \leq d(\lambda)
+ 2^{-r}$.  To see this note that
\begin{align*}
\Phi^+_r(d)(\lambda) &+ \Phi^-_r(d)(\lambda) \\
&= \nu(\CantorSet_{A^+}) + \nu(\CantorSet_{A^-}) \\
&\leq \nu(S^1[d] \cap X) + \nu(S^1[d] - X) + 2^{-r} \\
&= \nu(S^1[d]) + 2^{-r} \\
&\leq d(\lambda) + 2^{-r}
\end{align*}
by Lemma~\ref{G.3}.
\end{enumerate}

We have now shown that $\Phi$ is an all-$\nu$-measurement of $X$,
whence $X$ is all-$\nu$-measurable.
\end{proofof}

In the classical case, two definitions of $\nu$-measurability are
frequently used.  (See Royden \cite{Royd88} for example.)
The definition we have used above is known as the
Carath\'{e}odory definition.  The other definition frequently used is
the Lebesgue definition, which is stated in terms of equality of inner
and outer measures.  These two definitions are known to be equivalent
in the classical case.  Our definition of $\Delta$-$\nu$-measurability
in terms of $\Delta$-computable splitting operators is motivated by
the classical Carath\'{e}odory definition, and the proof of
Theorem~\ref{M.6} articulates this motivation more precisely.
Corollary~\ref{Z5B} below suggests that resource-bounded
measurability in complexity classes does not have
an equivalent definition in the style of Lebesgue.

We now note that the $\Delta$-{measures} of $\Delta$-{measurable} sets
must themselves be $\Delta$-{computable}.

\begin{theorem}
\label{M.7}
Let $X \subseteq \CantorSet$.
\begin{enumerate}
\item[1.] If $X$ is $\nu$-measurable in $R(\Delta)$, then $\nu(X \mid
R(\Delta))$ is $\Delta$-computable.
\item[2.] If $X$ is $\Delta$-$\nu$-measurable, then $\nu_\Delta(X)$ is
$\Delta$-computable.
\end{enumerate}
\end{theorem}
\begin{proofof}[Theorem~\ref{M.7}]
It suffices to prove 1, since 2 then follows by part 2 of
Lemma~\ref{M.5}.

Assume that $X$ is $\nu$-measurable in $R(\Delta)$ with
the $\nu$-measurement $\Phi$
as witness.  Let $\widehat{\Phi}^+$ be a $\Delta$-computation of
$\Phi^+$, and define $\func{f}{\N}{\Q}$ by
\[
f(r) = \widehat{\Phi}^+_{r+1, r+1}(\one)(\lambda).
\]
Then $f$ is $\Delta$-computable and, for all $r \in \N$,
Lemma~\ref{M.4} tells us that
\begin{align*}
&|f(r) - \nu(X \mid R(\Delta))| \\
&\leq |f(r) - \Phi^+_{r+1}(\one)(\lambda)| +
|\Phi^+_{r+1}(\one)(\lambda) - \nu(X \mid R(\Delta))| \\
&\leq 2^{-(r+1)} + 2^{-(r+1)} \\
&= 2^{-r},
\end{align*}
so $f$ is a $\Delta$-computation of $\nu(X \mid R(\Delta))$.
\end{proofof}

Our next step is to show that cylinders are $\Delta$-$\nu$-measurable,
provided that $\nu$ itself is a $\Delta$-probability measure, as
defined in section~\ref{Martingales}.

\begin{lemma}
\label{M.8}
If $\nu$ is a $\Delta$-probability measure on $\CantorSet$, then for
each $w \in \strings$, the cylinder $\CantorSet_w$ is
$\Delta$-$\nu$-measurable, with $\nu_\Delta( \CantorSet_w) = \nu(w)$.
\end{lemma}
\begin{proofof}[Lemma~\ref{M.8}]
Assume the hypothesis, and let $w \in \strings$.  We have two cases.
\begin{enumerate}
\item[I:] $\nu(w)=0$.  (Note that $w \neq \lambda$ in this
case because $\nu(\lambda) = 1$.) Then define
\[
\func{\Phi}{\N \times {\cal D}_\nu}{{\cal D}_\nu \times {\cal D}_\nu}
\]
as follows.  For each $r \in \N$, $d \in {\cal D}_\nu$, and $v \in
\strings$, set
\begin{align*}
&\Phi^+_r(d)(v) =
\begin{cases}
1 & \text{if $w \sqsubseteq v$} \\
0 & \text{otherwise,}
\end{cases} \\
& \Phi^-_r(d)(v) = d(v).
\end{align*}
It is clear that $\Phi$ is $\Delta$-computable and that $\Phi^+_r(d)$,
$\Phi^-_r(d) \in {\cal D}_\nu$ for all $r \in \N$ and $d \in {\cal
D}_\nu$. Also, for all $r \in \N$ and $d \in {\cal D}_\nu$,
the following conditions hold.
\begin{enumerate}
\item[(i)] $S^1[d] \cap \CantorSet_w \subseteq \CantorSet_w =
S^1[\Phi^+_r(d)]$.
\item[(ii)]$S^1[d] - \CantorSet_w \subseteq S^1[d] =
S^1[\Phi^-_r(d)]$.
\item[(iii)] $\Phi^+_r(d)(\lambda) + \Phi^-_r(d)(\lambda) =
d(\lambda)$.
\end{enumerate}
Thus $\Phi$ is a $\Delta$-$\nu$-measurement of $\CantorSet_w$, so
$\CantorSet_w$ is $\Delta$-$\nu$-measurable, with
\begin{align*}
\nu_\Delta(\CantorSet_w) &= \Phi^+_\infty \\
&= \inf_{r \in \N} \Phi^+_r(\one)(\lambda) \\
&= \Phi^+_0(\one)(\lambda) \\
&= 0 \\
&= \nu( w ).
\end{align*}
\item[II:] $\nu(w) > 0$.  In this case, define
\[
\func{\Phi}{{\cal D}_\nu}{{\cal D}_\nu \times {\cal D}_\nu}
\]
as follows.  For each $d \in {{\cal D}_\nu}$ and $v \in \strings$, set
\begin{align*}
&\Phi^+(d)(v) =
\begin{cases}
d(w)\nu(w \mid v) & \text{if $v \sqsubseteq w$} \\
d(v) & \text{if $w \sqsubseteq v$} \\
0 & \text{otherwise},
\end{cases} \\
&\Phi^-(d)(v) = d(v) - \Phi^+(d)(v).
\end{align*}
Then define
\[
\func{\Psi}{\N \times {\cal D}_\nu}{{\cal D}_\nu \times {\cal D}_\nu}
\]
by
\[
\Psi(r, d) = \Phi(\Lambda(d)),
\]
where $\Lambda$ is the functional given by the Regularity Lemma.  It is
clear that $\Psi$ is $\Delta$-computable.

Let $d \in {\cal D}_\nu$.  It is routine to check that $\Phi^+(d) \in
{\cal D}_\nu$.  Also, for $v \sqsubseteq w$,
\[
d(w)\nu(w) \leq \sum_{ |u| = |w| }
d(u)\nu(u) = d(v)\nu(v),
\]
so $d(w)\nu(w \mid v) \leq d(v)$.  It follows readily that
$\Phi^+(d)(v) \leq d(v)$ for all $v \in \strings$, whence $\Phi^-(d)
\in {\cal D}_\nu$.  This confirms that $\Psi$ maps $\N \times {\cal
D}_\nu$ into ${\cal D}_\nu \times {\cal D}_\nu$.

Now let $r \in \N$ and $d \in {\cal D}_\nu$.

To see that $S^1[d] \cap \CantorSet_w \subseteq S^1[\Psi^+_r(d)]$, let
$A \in S^1[d] \cap \CantorSet_w$.  Then $A \in S^1[\Lambda(d)] \cap
\CantorSet_w$, so there exists $n \geq |w|$ such that $w \sqsubseteq
A[0..n-1]$ and $\Lambda(d)(A[0..n-1]) \geq 1$. Then
\begin{align*}
\Psi^+_r(d)(A[0..n-1]) &= \Phi^+(\Lambda(d))(A[0..n-1]) \\
&= \Lambda(d)(A[0..n-1]) \\
&\geq 1,
\end{align*}
so $A \in S^1[\Psi^+_r(d)]$.

To see that $S^1[d] - \CantorSet_w \subseteq S^1[\Psi^-_r(d)]$, let $A
\in S^1[d] - \CantorSet_w$.  Then $A \in S^1[\Lambda(d)] -
\CantorSet_w$, so there exists $n \geq |w|$ such that
$\Lambda(d)(A[0..n-1]) \geq 1$ and neither of $w$, $A[0..n-1]$ is a
prefix of the other.  Then $\Phi^+(\Lambda(d))(A[0..n-1]) = 0$, so
\begin{align*}
\Psi^-_r(d)(A[0..n-1]) &= \Phi^-(\Lambda(d))(A[0..n-1]) \\
&= \Lambda(d)(A[0..n-1]) \\
&\geq 1,
\end{align*}
so $A \in S^1[\Psi^-_r(d)]$.

Finally, note that
\begin{align*}
\Psi^+_r(d)(\lambda) + \Psi^-_r(d)(\lambda) &=
\Phi^+(\Lambda(d))(\lambda) + \Phi^-(\Lambda(d))(\lambda) \\
&= \Lambda(d)(\lambda) \\
&= d(\lambda).
\end{align*}

We have now shown that $\Psi$ is a $\Delta$-$\nu$-measurement of
$\CantorSet_w$.  The $\Delta$-$\nu$-measure of $\CantorSet_w$ is then
\begin{align*}
\nu_\Delta( \CantorSet_w ) &= \Psi^+_\infty \\
&= \Phi^+(\Lambda(\one))(\lambda) \\
&= \Phi^+(\one)(\lambda) \\
&= \one(w)\nu(w \mid \lambda) \\
&= \nu(w).
\end{align*}
\end{enumerate}
\end{proofof}

Lemma~\ref{M.8} has important consequences for measure in
$R(\Delta)$, as shown by the following corollaries and example.

\begin{corollary}
\label{M.9}
If $\nu$ is a $\Delta$-probability measure on $\CantorSet$, then for
each $w \in \strings$, the cylinder $\CantorSet_w$ is $\nu$-measurable
in $R(\Delta)$, with $\nu(\CantorSet_w \mid R(\Delta)) = \nu(w)$.
\end{corollary}
\begin{proofof}[Corollary~\ref{M.9}]
This follows immediately from Lemma~\ref{M.8} and part 2 of
Lemma~\ref{M.5}.
\end{proofof}

\begin{corollary}
\label{M.10}
If $\nu$ is a $\Delta$-probability measure on $\CantorSet$, then
$R(\Delta)$ is $\nu$-measurable in $R(\Delta)$, with $\nu(R(\Delta)
\mid R(\Delta)) = 1$.
\end{corollary}
\begin{proofof}[Corollary~\ref{M.10}]
This follows immediately from Corollary~\ref{M.9} with $w =
\lambda$.
\end{proofof}

\begin{example}
\label{M.11}
Consider the uniform probability measure $\mu$ on $\CantorSet$ and
the set $\REC$, consisting of all decidable languages.  By
Corollary~\ref{M.10} (with $\nu = \mu$ and $\Delta =
\REC$), $\REC$ is $\mu$-measurable in $\REC$, with $\mu(\REC
\mid \REC) = 1$.  On the other hand, $\REC$ is countable, so $\REC$ is
(classically) $\mu$-measurable, with $\mu(\REC) = 0$.  Since
$\mu(\REC \mid \REC) \neq \mu(\REC)$, it follows by
Lemma~\ref{M.7} and parts 1 and 2 of Lemma~\ref{M.5} that
$\REC$ is not rec-$\mu$-measurable.  This example shows that the
converses of parts 1 and 2 of Lemma~\ref{M.5} do not hold.
\end{example}

The rest of this section is devoted to showing that resource-bounded
measurability and
measure are well-behaved with respect to set-theoretic operations.

\begin{definition*}
Let $R \subseteq \CantorSet$.  An {\it algebra on} $R$ is a collection
${\cal F}$ of subsets of $\CantorSet$ with the following three
properties.
\begin{enumerate}
\item[(i)] $R \in {\cal F}$.
\item[(ii)] If $X \in {\cal F}$, then $X^{\rm c} \in {\cal F}$.
\item[(iii)] If $X, Y \in {\cal F}$, then $X \cup Y \in {\cal F}$.
\end{enumerate}

If ${\cal F}$ is an algebra on $R$, then a {\it subalgebra of} ${\cal
F}$ {\it on } $R$ is a set ${\cal E} \subseteq {\cal F}$ that is also
an algebra on $R$.
\end{definition*}

In this paper, we only use the above definition in cases where $R =
\CantorSet$ or $R = R(\Delta)$.  In the latter case, our
terminology is somewhat nonstandard, in that we allow an algebra on
$R(\Delta)$ to have elements that are not themselves subsets of $R(\Delta)$.
Our terminology is in accordance with past usage in resource-bounded
measure, where it is often convenient to speak of ``the measure of $X$
in $R(\Delta)$''  (e.g., the measure of P/Poly in ESPACE) in cases
where $X$ is not a subset of $R(\Delta)$.  In any case, if one so
prefers, it is a straightforward matter to define the notion of an
{\it algebra in} $R(\Delta)$, and to adapt the results here to this
notion.

\begin{theorem}
\label{M.12}
The set ${\cal F}^\nu_{R(\Delta)}$, consisting of all sets that are
$\nu$-measurable in $R(\Delta)$, is an algebra on $R(\Delta)$.  For
$X, Y \in {\cal F}^\nu_{R(\Delta)}$, we have
\[
\nu(X^{\rm c} \mid R(\Delta) ) = 1 - \nu(X \mid R(\Delta))
\]
and
\[
\nu(X \cup Y \mid R(\Delta)) = \nu(X \mid R(\Delta)) + \nu(Y \mid
R(\Delta)) - \nu(X \cap Y \mid R(\Delta)).
\]
\end{theorem}
\begin{proofof}[Theorem~\ref{M.12}]
By Corollary~\ref{M.10}, $R(\Delta) \in {\cal
F}^\nu_{R(\Delta)}$.

Assume that $X \in {\cal F}^\nu_{R(\Delta)}$.  Then there is a
$\nu$-measurement $\Phi$ of $X$ in $R(\Delta)$.  Define
\[
\func{\Psi}{\N \times {\cal D}_\nu}{{\cal D}_\nu \times {\cal D}_\nu}
\]
\begin{align*}
\Psi^+_r(d) &= \Phi^-_r(d), \\
\Psi^-_r(d) &= \Phi^+_r(d)
\end{align*}
It is easily verified that $\Psi$ is a $\nu$-measurement of $X^{\rm
c}$ in $R(\Delta)$, so $X^{\rm c} \in {\cal F}^\nu_{R(\Delta)}$ and,
by Lemma~\ref{M.2},
\begin{align*}
\nu(X^{\rm c} \mid R(\Delta)) &= \Psi^+_\infty = \{\Phi^-_\infty \\
&= 1 - \Phi^+_\infty \\
&= 1 - \nu(X \mid R(\Delta)).
\end{align*}

Now assume that $X, Y \in {\cal F}^\nu_{R(\Delta)}$.  Let $\Phi$ and
$\Psi$ be $\nu$-measurements of $X$ and $Y$, respectively, in
$R(\Delta)$.  For each $a, b \in \{+, - \}$, define the functional
\[
\func{\Theta[ab]}{\N \times {\cal D}_\nu}{{\cal D}_\nu}
\]
by
\[
\Theta[ab]_r(d) = \Psi^b_{r+2}( \Phi^a_{r+1}(d)).
\]
It is clear that each of the four functionals $\Theta[ab]$ so defined
is $\Delta$-computable.  For all $r \in \N$ and $d \in {\cal D}_\nu$,
our choice of $\Phi$ and $\Psi$ gives us the inclusions
\begin{align*}
R(\Delta) \cap (X \cap Y) \cap S^1[d] & \subseteq S^1[\Theta[++]_r(d)],\\
R(\Delta) \cap (X - Y) \cap S^1[d] & \subseteq S^1[\Theta[+-]_r(d)], \\
R(\Delta) \cap (Y - X) \cap S^1[d] & \subseteq S^1[\Theta[-+]_r(d)], \\
R(\Delta) \cap {(X \cup Y)}^{\rm c} \cap S^1[d] & \subseteq
S^1[\Theta[--]_r(d)]
\end{align*}
and the bound
\begin{align*}
\Theta[++]_r(d)(\lambda) + \Theta[+-]_r(d)(\lambda) &+
\Theta[-+]_r(d)(\lambda) + \Theta[--]_r(d)(\lambda) \\
& \leq \Phi^+_{r + 1}(d)(\lambda) + \Phi^-_{r + 1}(d)(\lambda) +
2^{-(r + 1)} \\
&\leq d(\lambda) + 2^{-r}.
\end{align*}
It follows that the functionals
\begin{align*}
\Theta[X] &= (\Theta[++] + \Theta[+-], \; \Theta[-+]+\Theta[--]), \\
\Theta[Y] &= (\Theta[++] + \Theta[-+],\; \Theta[+-]+\Theta[--]), \\
\Theta[X \cap Y] &= (\Theta[++],\; \Theta[+-] + \Theta[-+] + \Theta[--]),
\\
\Theta[X \cup Y] &= (\Theta[++]+\Theta[+-]+\Theta[-+],\; \Theta[--] )
\end{align*}
are $\nu$-measurements of $X, Y, X \cap Y, X \cup Y$, respectively, in
$R(\Delta)$.  Thus $X \cap Y$ and $X \cup Y$ are $\nu$-measurable in
$R(\Delta)$, with
\begin{align*}
\nu(X \cup Y \mid R(\Delta)) &+ \nu(X \cap Y \mid R(\Delta)) \\
&= \lim_{r \rightarrow \infty} \Theta[X \cup Y]^+_r(\one)(\lambda) +
\lim_{r \rightarrow \infty} \Theta[X \cap Y]^+_r(\one)(\lambda) \\
&= \sum_{a, b \in \{+, - \}} \lim_{r \rightarrow \infty}
\Theta[ab]_r(\one)(\lambda) \\
&= \lim_{r \rightarrow \infty} \Theta[X]^+_r(\one)(\lambda) + \lim_{r
\rightarrow \infty} \Theta[Y]^+_r(\one)(\lambda) \\
&= \nu(X \mid R(\Delta)) + \nu( Y \mid R(\Delta)).
\end{align*}
\end{proofof}

The following immediate consequence of Theorem~\ref{M.12} says
that measure in $R(\Delta)$ is additive and monotone.

\begin{corollary}
\label{M.13}
Let $X, Y \in {\cal F}^\nu_{R(\Delta)}$.
\begin{enumerate}
\item[1.] If $X \cap Y \cap R(\Delta) = \emptyset$, then
\[
\nu(X \cup Y \mid R(\Delta)) = \nu(X \mid R(\Delta)) + \nu(Y \mid
R(\Delta)).
\]
\item[2.] If $X \cap R(\Delta) \subseteq Y$, then $\nu(X \mid R(\Delta))\leq \nu(Y
\mid R(\Delta))$.
\end{enumerate}
\end{corollary}

The following analogs of Theorem~\ref{M.12} and its corollary are
proven in similar fashion.

\begin{theorem}
\label{M.14}
The set ${\cal F}^\nu_\Delta$, consisting of all sets that are
$\Delta$-$\nu$-measurable, is an algebra over $\CantorSet$.  For $X, Y
\in {\cal F}^\nu_\Delta$, we have
\[
\nu_\Delta(X^{\rm c}) = 1 - \nu_\Delta(X)
\]
and
\[
\nu_{\Delta}(X \cup Y) = \nu_\Delta(X) + \nu_\Delta(Y) - \nu_\Delta(X
\cap Y).
\]
\end{theorem}

\begin{corollary}
\label{M.15}
Let $X, Y \in {\cal F}^\nu_\Delta$.
\begin{enumerate}
\item[1.] If $X \cap Y = \emptyset$, then
\[
\nu_\Delta(X \cup Y) = \nu_\Delta(X) = \nu_\Delta(Y).
\]
\item[2.] If $X \subseteq Y$, then $\nu_\Delta(X) \leq \nu_\Delta(Y)$.
\end{enumerate}
\end{corollary}

Note that by Example~\ref{M.11}, ${\cal F}^\mu_{\rm rec}
\varsubsetneqq {\cal F}^\mu_{\REC}$, and ${\cal F}^\mu_{\rm
rec}$ is not an algebra over $\REC$.

An important property of classical measurability is its completeness,
which means that all subsets of measure-$0$ sets are measurable.  We
now show that resource-bounded measurability also enjoys this property.

\begin{definition*}
Let ${\cal F}$ be a subalgebra of ${\cal F}^\nu_{R(\Delta)}$ on
$R(\Delta)$.  Then ${\cal F}$ is $\nu$-{\it{complete on}} $R(\Delta)$
if, for all $X, Y \subseteq \CantorSet$, if $X \subseteq Y \in {\cal
F}$ and $\nu(Y \mid R(\Delta)) = 0$, then $X \in {\cal F}$.
\end{definition*}

\begin{definition*}
Let ${\cal F}$ be a subalgebra of ${\cal F}^\nu_\Delta$ on
$\CantorSet$. Then ${\cal F}$ is $\Delta$-$\nu$-{\it{complete}} if,
for all $X, Y \subseteq \CantorSet$, if $X \subseteq Y \in {\cal F}$
and $\nu_\Delta(Y) = 0$, then $X \in {\cal F}$.
\end{definition*}

\begin{theorem}
\label{M.16}
\begin{enumerate}
\item[1.] The algebra ${\cal F}^\nu_{R(\Delta)}$ is $\nu$-complete on
$R(\Delta)$.
\item[2.] The algebra ${\cal F}^\nu_\Delta$ is
$\Delta$-$\nu$-complete.
\end{enumerate}
\end{theorem}
\begin{proofof}[Theorem~\ref{M.16}]
To prove part 1, assume that $X \subseteq Y \in {\cal F}$ and $\nu(Y
\mid R(\Delta)) = 0$.  Let $\Psi$ be a $\nu$-measurement of $Y$ in
$R(\Delta)$.  Define
\[
\func{\Phi}{\N \times {\cal D}_\nu}{{\cal D}_\nu \times {\cal D}_\nu}
\]
\[
\Phi_r(d) = (\Psi^+_r(\one), d).
\]
It is easy to check that $\Phi$ is a $\nu$-measurement of $X$ in
$R(\Delta)$, whence $X \in {\cal F}^\nu_{R(\Delta)}$.

The proof of part 2 is identical.
\end{proofof}

The rest of this section concerns infinitary unions and intersections
of resource-bounded measurable sets.  Classical measure is countably
additive, which means that the union of a countable collection of
disjoint measurable sets is measurable, and the measure of the union
is the sum of the measures of the disjoint measurable sets.  However,
this property does not hold for arbitrary resource bounds $\Delta$.
For example, if $\Delta = {\rm p}$ or $\Delta = {\rm rec}$, then
$R(\Delta)$ can be written as a countable union of singleton sets,
each of which has $\Delta$-$\mu$-measure 0 \cite{Lutz:AEHNC}, hence
$\mu$-measure 0 in $R(\Delta)$.  The union of these singletons, $R(\Delta)$,
has $\mu$-measure 1 in $R(\Delta)$ -- exceeding the sum of the
$\mu$-measures of the singletons -- and is not
$\Delta$-$\mu$-measurable at all.  Thus any resource-bounded
analog of countable additivity must involve a restricted class of
countable unions.

The obvious approach is to restrict attention to countable sequences
of sets that are not only individually measurable, but for which there
is a uniform witness of their measurability.  This leads to the
following definitions.

\begin{definition*}
Let ${\cal F}$ be a subalgebra of ${\cal F}^\nu_{R(\Delta)}$ on
$R(\Delta)$.  A $\Delta$-{\it{sequence in}} ${\cal F}$ is a sequence
$(X_k \mid k \in \N)$ of sets $X_k \in {\cal F}$ for which there
exists a $\Delta$-computable functional
\[
\func{\Phi}{\N \times \N \times {\cal D}_\nu}{{\cal D}_\nu \times
{\cal D}_\nu}
\]
such that, for each $k \in \N$, $\Phi_k$ is a $\nu$-measurement of
$X_k$ in $R(\Delta)$.
\end{definition*}

\begin{definition*}
Let ${\cal F}$ be a subalgebra of ${\cal F}^\nu_\Delta$ on
$\CantorSet$.  A $\Delta$-{\it{sequence in}} ${\cal F}$ is a sequence
$(X_k \mid k \in \N)$ of sets $X_k \in {\cal F}$ for which there
exists a $\Delta$-computable functional
\[
\func{\Phi}{\N \times \N \times {\cal D}_\nu}{{\cal D}_\nu \times
{\cal D}_\nu}
\]
such that, for each $k \in \N$, $\Phi_k$ is a
$\Delta$-$\nu$-measurement of $X_k$.
\end{definition*}

Although these definitions are useful for our purposes here,
the following example shows that they are not sufficient.

\begin{example}
\label{M.17}
Let $(M_i \mid i \in \N)$ be a standard enumeration of Turing
machines.  Using the standard enumeration $(s_j \mid j \in \N)$ of
$\strings$ and a standard pairing bijection $<, > : \N \times \N
\longrightarrow \N$, define a sequence $(X_k \mid k \in \N)$ as
follows.  For $k = <i, j> \in \N$, if $M_i(0^i)$ halts in exactly
$|s_j|$ steps, then let $X_k = \CantorSet_{0^i1s_j}$; otherwise, let
$X_k = \emptyset$.  Since the construction in the proof of
Lemma~\ref{M.8} is easily uniformized, it is not hard to show
that $(X_k \mid k \in \N)$ is a ${\rm p}$-sequence in ${\cal
F}^\mu_{\rm p}$.  Let $X = \bigcup_{k=0}^\infty X_k$.  Then the sets
$X_0, X_1, \ldots$ are disjoint, so
\[
\mu(X) = \sum_{k=0}^\infty \mu(X_k) = \frac{1}{2} \sum_{i \in
K} 2^{-i},
\]
where
\[
K = \set{i \in \N}{M_i(0^i) \text{ halts}}
\]
is the diagonal halting problem.  It is thus clear that $\mu(X)$
is Turing-equivalent to $K$, hence not computable.  (In fact,
$\mu(X)$ is a version of Chaitin's random real number $\Omega$
\cite{Chai75a}.)  It follows by Lemmas~\ref{M.5} and \ref{M.7}
that $X \not \in {\cal F}^\mu_{\rm rec}$.  We thus have a
${\rm p}$-sequence of disjoint elements of ${\cal F}^\mu_{\rm p}$ whose
union is not an element of ${\cal F}^\mu_{\rm p}$.
\end{example}

The set $X$ of Example~\ref{M.17} fails to be
rec-$\mu$-measurable because the sum $\sum^\infty_{k=0} \mu
(X_k)$ converges too slowly.  The following technical definition
prohibits such phenomena.

\begin{definition*}
A functional
\[
\func{\Phi}{\N \times \N \times {\cal D}_\nu}{{\cal D}_\nu \times
{\cal D}_\nu}
\]
is $\Delta$-{\it{modulated}} if the sequences $(\Phi^a_{k, r}(d)(w)
\mid k \in \N)$, for $a \in \{+, -\}$, $r \in \N$, $d \in {\cal
D}_\nu$, and $w \in \strings$, are uniformly $\Delta$-convergent.
Equivalently, $\Phi$ is $\Delta$-modulated if there is a
$\Delta$- computable functional
\[
\func{\Gamma}{\N \times \N \times {\cal D}_\nu \times \strings}{\N}
\]
such that, for all $a \in \{+, -\}$, $t, r \in \N$, $d \in {\cal
D}_\nu$, $w \in \strings$, and $k \geq \Gamma(t, r, d, w)$,
\[
\left| \Phi^a_{k, r}(d)(w) - \Phi^a_{\infty, r}(d)(w) \right| \leq
2^{-t}
\]
where the limit $\Phi^a_{\infty, r}(d)(w) = \lim_{k \rightarrow
\infty} \Phi^a_{k, r}(d)(w)$ is implicitly assumed to exist.
\end{definition*}

We now define the restricted infinitary unions and intersections with
which we deal.

\begin{definition*}
Let ${\cal F}$ be a subalgebra of ${\cal F}^\nu_{R(\Delta)}$ on
$R(\Delta)$.
\begin{enumerate}
\item[1.] A {\it union} $\Delta$-{\it{sequence in}} ${\cal F}$ is a
sequence $(X_k \mid k \in \N)$ of sets $X_k \in {\cal F}$ for which
there exists a $\Delta$-modulated functional $\Phi$ such that
$\Phi^+_{k,r}(d)(w)$ is nondecreasing in $k$, $\Phi^-_{k, r}(d)(w)$ is
nonincreasing in $k$, and $\Phi$ testifies that $\left(
\bigcup^{k-1}_{j=0} X_j \mid k \in \N\right)$ is a $\Delta$-sequence
in ${\cal F}$.
\item[2.] An {\it intersection} $\Delta$-{\it{sequence in}} ${\cal F}$
is a sequence $(X_k \mid k \in \N)$ of sets $X_k \in {\cal F}$ for
which there exists a $\Delta$-modulated functional $\Phi$ such that
$\Phi^+_{k, r}(d)(w)$ is nonincreasing in $k$, $\Phi^-_{k,r}(d)(w)$ is
nondecreasing in $k$, and $\Phi$ testifies that $\left(
\bigcap^{k-1}_{j=0} X_j \mid k \in \N \right)$ is a $\Delta$-sequence
in ${\cal F}$.
\item[3.] ${\cal F}$ is {\it closed under} $\Delta$-{\it{unions}} if
$\bigcup^\infty_{k=0} X_k \in {\cal F}$ for every union
$\Delta$-sequence $(X_k \mid k \in \N)$ in ${\cal F}$.
\item[4.] ${\cal F}$ is {\it closed under}
$\Delta$-{\it{intersections}} if $\bigcap_{k=0}^\infty X_k \in {\cal
F}$ for every intersection $\Delta$-sequence $(X_k \mid k \in \N)$ in
${\cal F}$.
\end{enumerate}
\end{definition*}

The motivation for these definitions will be discussed further in the
final version of this paper. We note in the next lemma that, when the
sets $X_k$ have resource-bounded measure 0, every $\Delta$-sequence is
already a union $\Delta$-sequence.

\begin{lemma}
\label{M.18}
Let ${\cal F}$ be a subalgebra of ${\cal F}^\nu_{R(\Delta)}$ on
$R(\Delta)$.  If $(X_k \mid k \in \N)$ is a $\Delta$-sequence in
${\cal F}$ and $\nu(X_k \mid R(\Delta)) = 0$ for all $k \in \N$, then
$(X_k \mid k \in \N)$ is a union $\Delta$-sequence in ${\cal F}$.
\end{lemma}
\begin{proofof}[Lemma~\ref{M.18}]
Assume the hypothesis, and let $\Phi$ be a witness that $(X_k \mid k
\in  \N)$ is a $\Delta$-sequence in ${\cal F}$.  Then the functional
\[
\func{\Psi}{\N \times \N \times {\cal D}_\nu}{{\cal D}_\nu \times
{\cal D}_\nu}
\]
\[
\Psi_{k,r}(d) = \left( \sum_{j=0}^k \Phi^+_{j, j+r+1}(\one), d\right)
\]
is easily seen to testify that $(X_k \mid k \in \N)$ is a union
$\Delta$-sequence in ${\cal F}$.
\end{proofof}

The following easy de Morgan law is useful.

\begin{lemma}
\label{M.19}
Let ${\cal F}$ be a subalgebra of ${\cal F}^\nu_{R(\Delta)}$ on
$R(\Delta)$.  Then a sequence $(X_k \mid k \in \N)$ is a union
$\Delta$-sequence in ${\cal F}$ if and only if the complemented
sequence $(X^{\rm c}_k \mid k \in \N)$ is an intersection
$\Delta$-sequence in ${\cal F}$.  Thus ${\cal F}$ is closed under
$\Delta$-unions if and only if ${\cal F}$ is closed under
$\Delta$-intersections.
\end{lemma}
\begin{proofof}[Lemma~\ref{M.19}]
It is clear that $(\Phi^+, \Phi^-)$ testifies that $(X_k \mid k \in
\N)$ is a union $\Delta$-sequence in ${\cal F}$ if and only if
$(\Phi^-, \Phi^+)$ testifies that $(X^{\rm c}_k \mid k \in \N)$ is an
intersection $\Delta$-sequence in ${\cal F}$.
\end{proofof}

We now show that measure in $R(\Delta)$ is well-behaved with respect
to $\Delta$-unions and $\Delta$-intersections.

\begin{theorem}
\label{M.20}
\begin{enumerate}
\item[1.] ${\cal F}^\nu_{R(\Delta)}$ is closed under $\Delta$-unions
and $\Delta$-intersections.
\item[2.] If $(X_k \mid k \in \N)$ is a union $\Delta$-sequence in
${\cal F}^\nu_{R(\Delta)}$, then
\[
\nu(\cup_{k=0}^\infty X_k \mid R(\Delta)) \leq \sum_{k=0}^\infty
\nu(X_k \mid R(\Delta)),
\]
with equality if the sets $X_0, X_1, \ldots$ are disjoint.
\item[3.] If $(X_k \mid k \in \N)$ is a union $\Delta$-sequence in
${\cal F}^\nu_{R(\Delta)}$ with each $X_k \subseteq X_{k+1}$, then
\[
\nu( \cup_{k=0}^\infty X_k \mid R(\Delta)) = \lim_{k \rightarrow
\infty} \nu(X_k \mid R(\Delta)).
\]
\item[4.] If $(X_k \mid k \in \N)$ is an intersection
$\Delta$-sequence in ${\cal F}^\nu_{R(\Delta)}$ with each $X_{k+1}
\subseteq X_k$, then
\[
\nu\left(\cap_{k=0}^\infty X_k \mid R(\Delta)\right) = \lim_{k
\rightarrow \infty} \nu \left(X_k \mid R(\Delta) \right).
\]
\end{enumerate}
\end{theorem}
\begin{proofof}[Theorem~\ref{M.20}]
Let $(X_k \mid k \in \N)$ be a union $\Delta$-sequence in ${\cal
F}^\nu_{R(\Delta)}$, with the functional $\Phi$ as witness, and let $X
\ \bigcup_{k=0}^\infty X_k$.  Fix a functional $\Gamma$ testifying
that $\Phi$ is $\Delta$-modulated, and define a functional
\[
\func{\Theta}{\N \times {\cal D}_\nu}{{\cal D}_\nu \times {\cal
D}_\nu}
\]
by
\[
\Theta^+_r(d)(w) = \Phi^+_{\infty, r+1}(d)(w)
\]
and
\[
\Theta^-_r(d)(w) = \Phi^-_{m, r+1}(d)(w),
\]
where $m = \Gamma( r+1, r+1, d, \lambda)$.  By our choice of $\Phi$
and $\Gamma$, $\Theta$ is a well-defined, $\Delta$-computable
functional mapping $\N \times {\cal D}_\nu$ into ${\cal D}_\nu \times
{\cal D}_\nu$.  To see that $\Theta$ is a $\nu$-measurement of $X$ in
$R(\Delta)$, let $r \in \N$ and $d \in {\cal D}_\nu$, and note the
following three things.
\begin{enumerate}
\item[1.] $R(\Delta) \cap X \cap S^1[d] \subseteq
S^1[\Theta^+_r(d)]$.  To see this, let $A \in R(\Delta) \cap X \cap
S^1[d]$.  Then there exists $k \in \N$ such that
\begin{align*}
A \in R(\Delta) \cap \left( \cup_{j=0}^{k-1} X_j \right) \cap
S^1[d] &\subseteq S^1[\Phi^+_{k, r+1}(d)] \\
&\subseteq S^1[\Phi^+_{\infty, r+1}(d)] \\
&= S^1[\Theta^+_r(d)].
\end{align*}
\item[2.] $R(\Delta) \cap X^{\rm c} \cap S^1[d] \subseteq
S^1[\Theta^-_r(d)]$.  To see this, let $A \in R(\Delta) \cap X^{\rm c}
\cap S^1[d]$.  Then, writing $m = \Gamma(r+1, r+1, d, \lambda)$, we
have
\begin{align*}
A \in &R(\Delta) \cap \left( \cup_{j=0}^{m-1} X_j\right)^{\rm c}
\cap S^1[d] \\
&\subseteq S^1[\Phi^-_{m, r+1}(d)] \\
&= S^1[\Theta^-_r(d)].
\end{align*}
\item[3.] $\Theta^+_r(d)(\lambda) + \Theta^-_r(d)(\lambda) \leq
d(\lambda) + 2^{-r}$.  To see this, let $m = \Gamma(r+1, r+1, d,
\lambda)$.  Then
\begin{align*}
\Theta^+_r(d)(\lambda) &= \Phi^+_{\infty, r+1}(d)(\lambda) \\
&\leq \Phi^+_{m, r+1}(d)(\lambda) + 2^{-(r+1)},
\end{align*}
so
\begin{align*}
\Theta^+_r(d)(\lambda) &+ \Theta^-_r(d)(\lambda) \\
&\leq \Phi^+_{m, r+1}(d)(\lambda) + \Phi^-_{m, r+1}(d)(\lambda) +
2^{-(r+1)} \\
&\leq d(\lambda) + 2^{-r}.
\end{align*}
\end{enumerate}

We have now shown that $\Theta$ is a $\nu$-measurement of $X$ in
$R(\Delta)$, whence $X \in {\cal F}^\nu_{R(\Delta)}$.  This, together
with Lemma~\ref{M.19}, proves part 1 of the theorem.

Since $\Theta$ is a $\nu$-measurement of $X$ in $R(\Delta)$,
Theorem~\ref{M.12} and its corollary (applied inductively) tell
us that
\begin{align*}
\nu\left( \bigcup_{k=0}^\infty X_k \mid R(\Delta)\right) &= \lim_{r
\rightarrow \infty} \Theta^+_r(\one)(\lambda) \\
&= \lim_{r \rightarrow \infty} \lim_{k \rightarrow \infty} \Phi^+_{k,
r+1}(\one)(\lambda) \\
&= \lim_{k \rightarrow \infty} \lim_{r \rightarrow \infty} \Phi^+_{k,
r+1}(\one)(\lambda) \\
&= \lim_{k \rightarrow \infty} \nu \left( \bigcup_{j=0}^{k-1} X_j \mid
R(\Delta) \right) \\
&\leq \lim_{k \rightarrow \infty} \sum_{j=0}^{k-1} \nu(X_j \mid
R(\Delta)) \\
&= \sum_{k=0}^\infty \nu(X_k \mid R(\Delta)),
\end{align*}
with equality if the sets $X_0, X_1, \ldots$ are disjoint.  This
proves part 2 of the theorem.

To prove part 3 of the theorem, let $(X_k \mid k \in \N)$ be a union
$\Delta$-sequence in ${\cal F}^\nu_{R(\Delta)}$ with each $X_k
\subseteq X_{k+1}$.  Let $Y_0 = X_0$ and, for each $k \in \N$, let
$Y_{k+1} = X_{k+1} - X_k$.  By uniformly applying the technique of the
proof of Theorem~\ref{M.12} to a functional testifying that
$(X_k \mid k \in \N)$ is a union $\Delta$-sequence in ${\cal
F}^\nu_{R(\Delta)}$, it is routine to obtain a functional testifying
that $(Y_k \mid k \in \N)$ is also a union $\Delta$-sequence in ${\cal
F}^\nu_{R(\Delta)}$.  Since the sets $Y_0, Y_1, \ldots$ are disjoint
and $X_k = \bigcup_{j=0}^k Y_j$ for each $k \in \N$, it follows by
part 2 of this theorem and Theorem~\ref{M.12} that
\begin{align*}
\nu\left(\bigcup_{k=0}^\infty X_k \mid R(\Delta)\right) &= \nu\left(
\bigcup_{k=0}^\infty Y_k \mid R(\Delta)\right) \\
&= \sum_{k=0}^\infty \nu(Y_k \mid R(\Delta)) \\
&= \lim_{k \rightarrow \infty} \sum_{j=0}^{k} \nu(Y_j \mid
R(\Delta)) \\
&= \lim_{k \rightarrow \infty} \nu\left( \bigcup_{j=0}^{k} Y_j \mid
R(\Delta)\right) \\
&= \lim_{k \rightarrow \infty} \nu( X_k \mid R(\Delta)).
\end{align*}

To prove part 4 of the theorem, let $(X_k \mid k \in \N)$ be an
intersection $\Delta$-sequence in ${\cal F}^\nu_{R(\Delta)}$ with each
$X_{k+1} \subseteq X_k$.  By part 1 of  this theorem,
$\bigcap_{k=0}^\infty X_k \in {\cal F}^\nu_{R(\Delta)}$, so by
Theorem~\ref{M.12},
\begin{align*}
\nu \left( \bigcap_{k=0}^\infty X_k \mid R(\Delta)\right) &= 1 -
\nu\left( \left( \bigcap_{k=0}^\infty X_k\right)^{\rm c} \mid
R(\Delta) \right) \\
&= 1 - \nu\left( \bigcup_{k=0}^\infty X_k^{\rm c} \mid R(\Delta) \right).
\end{align*}
By Lemma~\ref{M.19}, the complemented sequence $(X^{\rm c} \mid k
\in \N)$ is a union $\Delta$-sequence in ${\cal F}^\nu_{R(\Delta)}$.
It is clear that each $X^{\rm c}_k \subseteq X^{\rm c}_{k+1}$, so by
part 3 of this theorem and Theorem~\ref{M.12},
\begin{align*}
\nu\left(\bigcap_{k=0}^\infty X_k \mid R(\Delta)\right) &= 1 - \lim_{k
\rightarrow \infty} \nu(X^{\rm c}_k \mid R(\Delta)) \\
&= 1 - \lim_{k \rightarrow \infty} (1 - \nu(X_k \mid R(\Delta))) \\
&= \lim_{k \rightarrow \infty} \nu( X_k \mid R(\Delta)).
\end{align*}
\end{proofof}

\begin{corollary}
\label{M.21}
If $(X_k \mid k \in \N)$ is a $\Delta$-sequence of sets, each of which
has $\nu$-measure $0$ in $R(\Delta)$, then $\nu(\bigcup_{k=0}^\infty
X_k \mid R(\Delta) ) = 0$.
\end{corollary}
\begin{proofof}[Corollary~\ref{M.21}]
This follows immediately from Lemma~\ref{M.19} and
Theorem~\ref{M.20}.
\end{proofof}

Our discussion of union $\Delta$-sequences and intersection
$\Delta$-sequences has been confined to the algebra ${\cal
F}^\nu_{R(\Delta)}$.  It is a routine matter to develop the analogous
definitions and results for the algebra ${\cal F}^\nu_\Delta$.

\section{Measure Zero}
\label{MZ}

In this section we prove a resource-bounded generalization of the
classical Kolmogorov zero-one law that holds in complexity classes,
and we develop some useful characterizations of sets of
resource-bounded measure 0.  We begin our development with two simple
consequences of resource-bounded measurability.
Throughout this section, $\Delta$ is a resource bound, as specified in
section~\ref{Prelim}, and $\nu$ is a probability measure on
$\CantorSet$, as defined in section~\ref{Martingales}.

\begin{definition*}
Let $X \in \CantorSet$.  A $\nu$-{\it{bicover of}} $X$ {\it in}
$R(\Delta)$ is an ordered pair $(d^+, d^-)$ of $\Delta$-computable
functions
\[
\func{d^+, d^-}{\N \times \strings}{[0, \infty)}
\]
that satisfy the following conditions for all $r \in \N$.
\begin{enumerate}
\item[(i)] $d^+_r$ and $d^-_r$ are $\nu$-martingales.
\item[(ii)] $R(\Delta) \cap X \subseteq S^1[d^+_r]$.
\item[(iii)] $R(\Delta) - X \subseteq S^1[d^-_r]$.
\item[(iv)] $d^+_r(\lambda) + d^-_r(\lambda) \leq 1 + 2^{-r}$.
\end{enumerate}
If $(d^+, d^-)$ is a $\nu$-bicover of $X$ in $R(\Delta)$, then we use
the notation
\[
d^+_\infty(\lambda) = \lim_{r \rightarrow \infty} d^+_r(\lambda).
\]
(This limit is shown to exist in Lemma~\ref{Z.1})
\end{definition*}

\begin{definition*}
Let $X \in \CantorSet$.  The {\it martingale outer} $\nu$-{\it{measure
of}} $X$ {\it in} $R(\Delta)$ is the infimum $\nu^\circ(X \mid
R(\Delta))$ of all $d(\lambda)$ for which $d$ is a
$\Delta$-$\nu$-martingale such that $X \cap R(\Delta) \subseteq
S^1[d]$.
\end{definition*}

Note that, for {\it every} set $X \subseteq \CantorSet$, $\nu^\circ(X
\mid R(\Delta))$ is defined and
\[
0 \leq \nu^\circ(X \mid R(\Delta)) \leq 1.
\]

\begin{lemma}
\label{Z.1}
Let $X \subseteq \CantorSet$, and consider the following three
conditions.
\begin{enumerate}
\item[(A)] $X$ is $\nu$-measurable in $R(\Delta)$.
\item[(B)] $X$ has a $\nu$-bicover in $R(\Delta)$.
\item[(C)] $\nu^\circ(X \mid R(\Delta)) + \nu^\circ(X^{\rm c} \mid
R(\Delta)) = 1$.
\end{enumerate}
Then $\text{(A) } \Longrightarrow \text{ (B) } \Longrightarrow \text{
(C)}$.  Moreover, we have the following.
\begin{enumerate}
\item[1.] If (A) holds and $(d^+, d^-)$ is any $\nu$-bicover of $X$ in
$R(\Delta)$, then $d^+_\infty(\lambda) = \nu(X|R(\Delta))$.
\item[2.] If (B) holds and $(d^+, d^-)$ is an $\nu$-bicover of $X$ in
$R(\Delta)$, then $d^+_\infty(\lambda) = \nu^\circ(X \mid R(\Delta))$.
\end{enumerate}
\end{lemma}
\begin{proofof}[Lemma~\ref{Z.1}]
First, assume that (A) holds, and let $\Phi$ be a $\nu$-measurement of
$X$ in $R(\Delta)$.  Then it is easy to check that $(\Phi^+(\one),
\Phi^-(\one))$ is a $\nu$-bicover of $X$ in $R(\Delta)$, whence (B)
holds.  Now let $(d^+, d^-)$ be any $\nu$-bicover of $X$ in
$R(\Delta)$.  Then, by the Measure Conservation Theorem (arguing as in
Lemma~\ref{M.1}), for all $r \in \N$,
\begin{align*}
1 &\leq \Phi^+_r(\one)(\lambda) + d^-_r(\lambda) \\
&\leq \Phi^+_r(\one)(\lambda) + (1 + 2^{-r} - d^+_r(\lambda)),
\end{align*}
so $d^+_r(\lambda) \leq \Phi^+_r(\one)(\lambda) + 2^{-r}$.  Similary,
$\Phi^+_r(\one)(\lambda) \leq d^+_r(\lambda) + 2^{-r}$, so
\[
\left| \Phi^+_r(\one)(\lambda) - d^+_r(\lambda) \right| \leq 2^{-r}.
\]
It follows immediately that
\[
d^+_\infty(\lambda) = \Phi^+_\infty = \nu(X \mid R(\Delta)).
\]

Next, assume that (B) holds, and let $(d^+, d^-)$ be a $\nu$-bicover
of $X$ in $R(\Delta)$.  The Measure Conservation Theorem (arguing
again as in Lemma 4.1) tells us that $\nu^\circ(X \mid R(\Delta)) +
\nu^\circ(X^{\rm c} \mid R(\Delta)) \geq 1$ in any case.  To see that
(C) holds, let $r \in \N$  be arbitrary.  Then $R(\Delta) \cap X
\subseteq S^1[d^+_r]$ and $R(\Delta) \cap X^{\rm c} \subseteq
S^1[d^-_r]$, so
\begin{align*}
\nu^\circ(X \mid R(\Delta)) &+ \nu^\circ(X^{\rm c} \mid R(\Delta)) \\
&\leq d^+_r(\lambda) + d^-_r(\lambda) \\
&\leq 1 + 2^{-r}.
\end{align*}
Since $r$ is arbitrary here, this confirms (C).  We have already noted
that $\nu^\circ(X \mid R(\Delta)) \leq d^+_r(\lambda)$ for all $r \in
\N$; it follows immediately that $\nu^\circ(X \mid R(\Delta)) \leq
d^+_\infty(\lambda)$.  To see that $d^+_\infty(\lambda) \leq
\nu^\circ(X \mid R(\Delta))$, let $d$ be an arbitrary
$\Delta$-$\nu$-martingale such that $X \cap R(\Delta) \subseteq
S^1[d]$, and let $r \in \N$ be arbitrary.  Then, by the Measure
Conservation Theorem (arguing yet again as in Lemma~\ref{M.1}),
\begin{align*}
1 &\leq d(\lambda) + d^-_r(\lambda) \\
&\leq d(\lambda) + (1 + 2^{-r} - d^+_r(\lambda)),
\end{align*}
so $d(\lambda) \geq d^+_r(\lambda) - 2^{-r}$.  Since $r$ is arbitrary,
it follows that $d(\lambda) \geq d^+_\infty(\lambda)$.  Since $d$ is
arbitrary, it follows in turn that $\nu^\circ(X \mid R(\Delta)) \geq
d^+_\infty(\lambda)$, completing the proof.
\end{proofof}

Before proceeding, we mention two examples and a lemma that further
illuminate Lemma~\ref{Z.1}

\begin{example}
\label{Z.2}
Consider the uniform probability measure $\mu$ on $\CantorSet$ and the
complexity class $R({\rm p}) = {\rm E}$.  For each $k \in \N$, let
\[
X_k = \DTIME(2^{(k+1)n}) - \DTIME(2^{kn}),
\]
and let $X = \bigcup_{k=0}^\infty X_{2k}$.  The diagonalization
technique used in the proof of the Measure Conservation Theorem can
readily be adapted to show that
\[
\nu^\circ(X \mid {\rm E}) = \mu^\circ( X^{\rm c} \mid {\rm E}) = 1.
\]
Thus condition (C) of Lemma~\ref{Z.1} is not satisfied by all $X
\subseteq \CantorSet$.
\end{example}

\begin{example}
\label{Z.3}
Again, consider the uniform probability measure $\mu$ on $\CantorSet$
and the complexity class $R({\rm p}) = {\rm E}$.  For each $k \in \N$, let
\[
X_k = \DTIME(2^{kn}) \cap \CantorSet_{0^k},
\]
and let $X = \bigcup_{k=0}^{\infty} X_k$.  We show in the full version
of this paper
that $X$ satisfies condition (C) of Lemma~\ref{Z.1}, with $\mu^\circ(X \mid
{\rm E}) = 0$, but that $X$ does not satisfy condition (B).  Thus, in
Lemma~\ref{Z.1}, condition (C) does not imply condition (B) -- or condition
(A) -- even in the measure $0$ case.
\end{example}
\begin{proofof}[Example~\ref{Z.3}]

To see that $X$ satisfies condition (C), let $r \in \N$ be arbitrary.  It is well-known \cite{Lutz:AEHNC} that $\mu_{\rm p}(\DTIME(2^{rn})) = 0$, so there is a ${\rm p}$-$\mu$-martingale $d^\prime$ such that $d^\prime(\lambda) \leq 2^{-(r+1)}$ and $\DTIME(2^{rn}) \subseteq S^1[d^\prime]$.  Define $\func{d^{\prime \prime}}{\strings}{[0, \infty)}$ by
\[
d^{\prime \prime}(w) =
\begin{cases}
2^{|w| - (r+1)} & \text{if $w \sqsubseteq 0^{r+1}$} \\
& \\
1 & \text{if $0^{r+1} \sqsubseteq w$} \\
& \\
0 & \text{otherwise.}
\end{cases}
\]
It is easy to check that $d^{\prime \prime}$ is a ${\rm
p}$-$\mu$-martingale with $d^{\prime \prime}(\lambda) = 2^{-(r+1)}$
and $S^1[d^{\prime \prime}] = \CantorSet_{0^{r+1}}$.  Thus, if we let
\[
d = d^\prime + d^{\prime \prime},
\]
then $d$ is a ${\rm p}$-$\mu$-martingale,
\[
d(\lambda) = d^\prime(\lambda) + d^{\prime \prime}(\lambda) \leq 2^{-r},
\]
and
\begin{align*}
X &= \left( \bigcup_{k=0}^r X_k \right) \cup \left( \bigcup_{k = r+1}^\infty X_k \right) \\
&\subseteq \DTIME(2^{rn}) \cup \CantorSet_{0^{r+1}} \\
&\subseteq S^1[d^\prime] \cup  S^1[d^{\prime \prime}] \\
&\subseteq S^1[d].
\end{align*}
Since $X \subseteq \E$, it follows that
\[
\nu^\circ(X \mid \E) \leq d(\lambda) \leq 2^{-r}.
\]
Since $r$ is arbitrary here, this shows that $\nu^\circ(X \mid \E) = 0$.  It follows directly from this that ( C ) holds.

To see that $X$ does not satisfy condition (B), let $\func{d^+}{\N
\times \strings}{[0, \infty)}$ be an arbitrary p-computable function
such that, for each $r \in \N$, $d^+_r$ is a p-$\mu$-martingale with
$d^+_r(\lambda) \leq 2^{-r}$.  Fix $k \in \N$ such that $d^+$ is
$n^k$-time computable.  Using the diagonalization technique of the
Measure Conservation Theorem, it is routine to show that $X_{k+1} \not
\subseteq S^1[d^+_{k+2}]$.  Since $X \subseteq \E$ and $d^+$ is
arbitrary, it follows that $X$ does not have a $\mu$-bicover in $\E$.
\end{proofof}

\begin{lemma}
\label{Z5A}
For every $A \in \REC$, there exist $X, Y \subseteq \CantorSet$ such
that $X$ and $Y$ have $\mu$-{bicovers} in $\E$ and $A \leq^{\rm
P}_{\rm T} \mu^{\circ}( X \cup Y)$.
\end{lemma}

\begin{corollary}
\label{Z5B}
There exist $X, Y \subseteq \CantorSet$ such that $X$ and $Y$ have
$\mu$-{bicovers} in $\E$ but $X \cup Y$ does not.
\end{corollary}

By Corollary~\ref{Z5B} and Theorem~\ref{M.12}, condition (B) of
Lemma~\ref{Z.1}  does not imply condition (A).  This appears to
suggest that, in contrast with the classical case, resource-bounded
measurability in complexity classes
does not admit a Lebesgue-style characterization in
terms of type-1 objects.  More seriously, Corollary~\ref{Z5B} says
that the sets having $\mu$-{bicovers} in $\E$ do not even form an
algebra on $\E$.  Note also that
condition (C) is noneffective, in the sense that there need be no
concrete witness to its being satisfied.  Thus the sets that satisfy
condition (C) cannot enjoy a closure property of the type established
for ${\cal F}^\nu_{R(\Delta)}$ in Theorem~\ref{M.20}.

We now turn to the zero-one law.

\begin{definition*}
A set $X \subseteq \CantorSet$ is a {\it tail set} if, for all $A, B
\in \CantorSet$, if $A \in X$ and the symmetric difference $(A-B) \cup (B-A)$
is finite, then $B \in X$.
\end{definition*}

Most subsets of $\CantorSet$ that are of interest in computational
complexity are tail sets.  If we are working with a coin-toss
probability measure on $\CantorSet$, i.e., a probability measure
$\mu^{\bvec}$ as defined in section 3, then the classical Kolmogorov
zero-one law \cite{Kolm33} says that every ${\bvec}$-measurable tail
set has ${\bvec}$-measure $0$ or ${\bvec}$-measure $1$.  (We
simplify terminology by writing ``${\bvec}$-measurable'' in place of
``$\mu^{\bvec}$-measurable,'' etc.)  That is, if $X \in {\cal
F}^{\bvec}$ is a tail set, then $\mu^{\bvec}(X) = 0$ or
$\mu^{\bvec}(X) = 1$.

By Lemma~\ref{M.5}, the Kolmogorov zero-one law trivially implies its own
generalization to the algebra ${\cal F}^{\bvec}_\Delta$.  That is, if
$X$ is a $\Delta$-${\bvec}$-measurable tail set, then
$\mu^{\bvec}_\Delta(X) = 0$ or $\mu^{\bvec}_\Delta(X) = 1$.  Our
objective here is to generalize the Kolmogorov zero-one law to the
algebra ${\cal F}^{\bvec}_{R(\Delta)}$.  This generalization does {\it not}
follow directly from the classical Kolmogorov zero-one law.  The main
part of our argument is the following lemma, which is a
resource-bounded zero-one law for the martingale outer
${\bvec}$-measure $\mu^{{\bvec} \circ}(X \mid R(\Delta))$ of an {\it
arbitrary} (not necessarily measurable) tail set $X$ in $R(\Delta)$.

\begin{lemma}
\label{Z.4}
If $X$ is a tail set and $\bvec$ is a $\Delta$-computable bias
sequence, then $\mu^{\bvec \circ}(X \mid R(\Delta)) = 0$ or $
\mu^{\bvec \circ}(X \mid R(\Delta)) = 1$.
\end{lemma}
\begin{proofof}[Lemma~\ref{Z.4}]
Assume the hypothesis, and let $\alpha = \mu^{\bvec \circ}(X \mid
R(\Delta))$.  It suffices to show that $\alpha \leq \alpha^2$.

Let $r \in \N$ be arbitrary, and fix a $\Delta$-$\bvec$-martingale $d$
such that $X \cap R(\Delta) \subseteq S^1[d]$ and $d(\lambda) \leq
\alpha + 2^{-(r+2)}$.  Without loss of generality, we can also assume
that $d$ is regular (by the Regularity Lemma) and that $d(\lambda)
\leq \mu^{\bvec}(S^1[d]) + 2^{-(r+3)}$ (doing finite surgery on $d$ to
acheive this, if necessary.)

For each $n \in \N$, let
\begin{align*}
I_n &= \set{w \in \strings}{d(w) < 1}, \\
J_n &= \set{w \in \{0, 1\}^n}{ d(w) \geq 1},
\end{align*}
and fix $m \in \N$ such that
\[
\sum_{w \in J_m} \mu^{\bvec}(w) \geq \mu^{\bvec}(S^1[d]) - 2^{-(r+3)}.
\]
Fix a string $u \in \{0, 1\}^m$ such that $d(u) \leq d(\lambda)$.
(Such a string $u$ must exist because $d$ is a martingale.)
Note that
\begin{align*}
\sum_{w \in I_m} d(w) \mu^{\bvec}(w) &= \sum_{w \in \{0, 1\}^m} d(w)
\mu^{\bvec}(w) - \sum_{w \in J_m} d(w) \mu^{\bvec}(w) \\
&= d(\lambda) - \sum_{w \in J_m} d(w) \mu^{\bvec}(w) \\
&\leq d(\lambda) - \sum_{w \in J_m} \mu^{\bvec}(w) \\
&\leq d(\lambda) - \mu^{\bvec}(S^1[d]) - 2^{-(r+3)} \\
& \leq 2^{-(r+2)}.
\end{align*}
and
\[
\sum_{w \in J_m} \mu^{\bvec}(w) \leq \mu^{\bvec}(S^1[d]) \leq
d(\lambda).
\]

Using the $\Delta$-$\bvec$-martingale $d$, the natural number $m$, the
sets $I_m$ and $J_m$, and the string $u$, define
$\func{d^\prime}{\strings}{[0, \infty)}$ as follows.  Let $w \in
\strings$.
\begin{enumerate}
\item[(i)]  If $|w| \geq m$ and $w[0..m-1] \in J_m$, then
\[
d^\prime(w) = d(u \ast w),
\]
where $u \ast w$ is the string obtained by substituting $u$ for the
first $m$ bits of $w$, i.e., $|u \ast w| = |w|$ and
\[
(u \ast w)[i] =
\begin{cases}
u[i] & \text{if $0 \leq i < m$} \\
w[i] & \text{if $m \leq i < |w|$}.
\end{cases}
\]
\item[(ii)]  If $|w| \geq m$ and $w[0..m-1] \in I_m$, then
\[
d^\prime(w) = d(w).
\]
\item[(iii)] If $|w| < m$, then $d^\prime(w)$ is defined from the
values of $d^\prime$ on $\{0, 1\}^m$ so that $d^\prime$ is a
$\bvec$-martingale.
\end{enumerate}
It is easy to check that $d^\prime$ is a $\Delta$-$\bvec$-martingale,
(The fact that we have a coin-toss probability measure is crucial here
in clause (i) of the definition of $d^\prime$.)

We now show that $X \cap R(\Delta) \subseteq S^1[d^\prime]$.  To see
this, let $A \in X \cap R(\Delta)$.  If $A[0..m-1] \in I_m$, then
$d^\prime(A[0..n-1]) = d(A[0..n-1])$ for all $n \geq m$.  Since $d$ is
regular and $A \in S^1[d]$, this implies that $A \in S^1[d^\prime]$.
On the other hand, if $A[0..m-1] \in J_m$, then for all $n \geq m$,
\begin{align*}
d^\prime(A[0..n-1]) &= d( u \ast (A[0..n-1])) \\
&= d((u \ast A)[0..n-1]),
\end{align*}
where $u \ast A \in \CantorSet$ is defined by
\[
(u \ast A)[i] =
\begin{cases}
u[i] & \text{if $0 \leq i < m$} \\
A[i] & \text{if $i \geq m$}.
\end{cases}
\]
Since $A \in X \cap R(\Delta)$ and $X$ is a tail set, we have
\[
u \ast A \in X \cap R(\Delta) \subseteq S^1[d].
\]
Since $d$ is regular, it now follows that $A \in S^1[d^\prime]$.  This
completes the demonstration that $X \cap R(\Delta) \subseteq
S^1[d^\prime]$.

Since $X \cap R(\Delta) \subseteq S^1[d^\prime]$, we now have
\begin{align*}
\alpha &= \mu^{\bvec \circ}(X \mid R(\Delta)) \leq d^\prime(\lambda) =
d^\prime(\lambda) \mu^{\bvec}(\lambda) \\
&= \sum_{w \in \{0, 1 \}^m} d^\prime(w) \mu^{\bvec}(w) \\
&\leq \sum_{w \in I_m} d^\prime(w) \mu^{\bvec}(w) + \sum_{w \in J_m}
d^\prime(w) \mu^{\bvec}(w) \\
&= \sum_{w \in J_m} d(w) \mu^{\bvec}(w) + d(u) \sum_{w \in J-m}
\mu^{\bvec}(w) \\
&\leq \sum_{w \in I_m} d(w) \mu^{\bvec}(w) + d(\lambda) \sum_{w \in
J_m} \mu^{\bvec}(w).
\end{align*}
Using the bounds we have already derived for the two sums in this last
expression, it follows that
\begin{align*}
\alpha &\leq2^{-(r+2)} + d(\lambda)^2 \\
&\leq 2^{-(r+2)} + (\alpha + 2^{-(r+2)})^2 \\
&\leq \alpha^2 + 2^{-r}.
\end{align*}
Since $r$ is arbitrary here, this establishes that $\alpha \leq
\alpha^2$, completing the proof.
\end{proofof}

Our resource-bounded generalization of the Kolmogorov zero-one law now
follows easily.

\begin{theorem}
\label{Z.5}
If $\bvec$ is a $\Delta$-computable bias sequence and $X$ is a tail
set that is $\bvec$-measurable in $R(\Delta)$, then $\mu^{\bvec}(X \mid
R(\Delta)) = 0$ or $\mu^{\bvec}(X \mid R(\Delta)) = 1$.
\end{theorem}
\begin{proofof}[Theorem~\ref{Z.5}]
Assume the hypothesis.  Then, by Lemma~\ref{Z.1}, $\mu^{\bvec}(X \mid
R(\Delta)) = \mu^{\bvec \circ}(X \mid R(\Delta))$, so the conclusion
follows immediately from Lemma~\ref{Z.4}
\end{proofof}

Theorem~\ref{Z.5}, like its classical counterpart, cannot be extended to arbitrary probability measures on $\CantorSet$.  For example, the set $X$ of all finite languages is a tail set, but if
\[
\nu(w) =
\begin{cases}
1 & \text{if $w = \lambda$} \\
& \\
\frac{1}{2} & \text{if $w \in \{0\}^+ \cup \{1\}^+$} \\
& \\
0 & \text{otherwise},
\end{cases}
\]
then $\nu(X) = \nu(X \mid R(\Delta)) = \frac{1}{2}$.

We conclude this paper with some useful characterizations of
resource-bounded measure $0$ sets.

The condition $\nu_\Delta(X \cap R(\Delta)) =
\alpha$ implies the condition $\nu(X \mid R(\Delta)) = \alpha$, but
the converse does {\it not} generally hold for $\alpha > 0$.  However,
when $\alpha = 0$, these two conditions are equivalent.

\begin{lemma}
\label{Z.6}
For any set $X \subseteq \CantorSet$, the following conditions are
equivalent.
\begin{enumerate}
\item[(1)] $\nu(X \mid R(\Delta)) = 0$.
\item[(2)] $\nu_\Delta(X \cap R(\Delta)) = 0$.
\end{enumerate}
\end{lemma}
\begin{proofof}[Lemma~\ref{Z.6}]
The fact that (2) implies (1) follows immediately from
Lemma~\ref{M.5}.  To see that (1) implies (2), assume (1).  Let $\Phi$
be a $\nu$-measurement of $X$ in $R(\Delta)$.  Then the functional
$\Psi$ defined by
\[
\Psi_r(d) = (\Phi^+_r(\one), d)
\]
is easily seen to be a $\Delta$-$\nu$-measurement of $X \cap
R(\Delta)$, so
\[
\nu_\Delta(X \cap R(\Delta)) = \Psi^+_\infty = \Phi^+_\infty = \nu(X
\mid R(\Delta)) = 0.
\]
\end{proofof}

We now give several useful characterizations of resource-bounded
measure $0$ sets.

\begin{definition*}
A $\Delta$-$\nu$-{\it{null cover}} of a set $X \subseteq \CantorSet$
is a $\Delta$-computable function $\func{d}{\N \times \strings}{[0,
\infty)}$ such that, for each $r \in \N$, $d_r$ is a $\nu$-martingale,
$X \subseteq S^1[d_r]$, and $d_r(\lambda) \leq 2^{-r}$.  A
$\Delta$-$\nu$-null cover $d$ is {\it regular} if the martingale $d_r$
is regular for each $r \in \N$.
\end{definition*}

Recall the martingale success sets $S^\infty[d]$ and $S^\infty_{\rm
str}[d]$ defined in section~\ref{Martingales}.

\begin{theorem}
\label{Z.7}
Let $X \in \CantorSet$, and let $\nu$ be a $\Delta$-probability
measure on $\CantorSet$.  The following conditions are equivalent.
\begin{enumerate}
\item[(1)] $\nu_\Delta(X) = 0$.
\item[(2)] $X$ has a $\Delta$-$\nu$-null cover.
\item[(3)] $X$ has a regular $\Delta$-$\nu$-null cover.
\item[(4)] There is a $\Delta$-$\nu$-martingale $d$ such that $X
\subseteq S^\infty_{\rm str}[d]$.
\item[(5)] There is a $\Delta$-$\nu$-martingale $d$ such that $X
\subseteq S^\infty[d]$.
\end{enumerate}
\end{theorem}
\begin{proofof}[Theorem~\ref{Z.7}]
To see that (1) implies (2), assume that $\nu_\Delta(X) = 0$, and let
$\Phi$ be a $\Delta$-$\nu$-measurement of $X$.  Then $\Phi^+(\one)$ is
easily seen to be a $\Delta$-$\nu$-null cover of $X$, so (2) holds.

It is clear by the Regularity Lemma that (2) implies (3).

To see that (3) implies (4), let $d^\prime$ be a regular
$\Delta$-$\nu$-null cover of $X$.  Define $\func{d}{\strings}{[0,
\infty)}$ by
\[
d(w) =
\begin{cases}
\sum_{r=0}^\infty d^\prime_r(w) & \text{if $\nu(w) > 0$} \\
|w| & \text{if $\nu(w) = 0$}
\end{cases}
\]
For all $w \in \strings$ such that $\nu(w) > 0$, the trivial
martingale inequality $d^\prime_r(w) \nu(w) \leq d^\prime_r(\lambda)$
assures us that
\begin{align*}
d(w) &= \sum_{r=0}^\infty d^\prime_r(w) \leq \frac{1}{\nu(w)}
\sum_{r=0}^\infty d^\prime(\lambda) \\
&\leq \frac{1}{\nu(w)} \sum_{r = 0}^\infty 2^{-r} = \frac{2}{\nu(w)}
\leq \infty,
\end{align*}
so $d$ is well-defined.  It is then easy to check that $d$ is a
$\Delta$-$\nu$-martingale.  To see that $X \subseteq S^\infty_{\rm
str}[d]$, let $A \in X$, and let $m \in \N$ be arbitrary.  We have two
cases.
\begin{enumerate}
\item[I:] There exists $w \sqsubseteq A$ such that $\nu(w)
= 0$.  Then, for all $n \geq \max \{ |w|, m\}$,
\[
d(A[0..n-1]) = n \geq m.
\]
\item[II:] $\nu(w) > 0$ for all $w \sqsubseteq A$.  For
each $0 \leq r < m$, fix $n_r \in \N$ such that
$d^\prime_r(A[0..n_r-1]) \geq 1$.  (Such $n_r$ exists because $A \in X
\subseteq S^1[d^\prime_r]$.)  Then, for all $n \geq \max \set{n_r}{0
\leq r < m}$, the fact that each $d^\prime_r$ is regular ensures that
\[
d(A[0..n-1]) \geq \sum_{r = 0}^{m-1} d^\prime_r(A[0..n-1]) \geq m.
\]
\end{enumerate}
In either case, we have shown that $d(A[0..n-1]) \geq m$ for all
sufficiently large $n$.  Since $m$ is arbitrary here, it follows that
$A \in S^\infty_{\rm str}[d]$, affirming (4).

It is trivial that (4) implies (5).

To see that (5) implies (1), let $d$ be a $\Delta$-$\nu$-martingale
such that $X \subseteq S^\infty[d]$.  For each $r \in \N$ and
$d^\prime \in {\cal D}_\nu$, let
\begin{align*}
\Phi^+_r(d^\prime) &= \frac{2^{-r}}{1 + d(\lambda)} \cdot d, \\
\Phi^-_r(d^\prime) &= d^\prime.
\end{align*}
It is easy to check that $\Phi = (\Phi^+, \Phi^-)$ is
$\Delta$-computable and that
\[
\func{\Phi}{\N \times {\cal D}_\nu}{{\cal D}_\nu \times {\cal D}_\nu}.
\]
To see that $\Phi$ is a $\Delta$-$\nu$-measurement of $X$, let $r \in
\N$ and $d^\prime \in {\cal D}_\nu$.  Three things are to be checked.
\begin{enumerate}
\item[(i)] $S^1[d^\prime] \cap X \subseteq S^1[\Phi^+_r(d^\prime)]$.
To see this, let $A \in S^1[d^\prime] \cap X$.  Then $A \in X
\subseteq S^\infty[d]$, so there exists $n \in \N$ such that
$d(A[0..n-1]) \geq 2^r \cdot (1 + d(\lambda))$.  Then
$\Phi^+_r(d^\prime)(A[0..n-1]) \geq 1$, so $A \in
S^1[\Phi^+_r(d^\prime)]$.
\item[(ii)] $S^1[d^\prime] -X \subseteq S^1[\Phi^-_r(d^\prime)]$.
This holds trivially.
\item[(iii)] $\Phi^+_r(d^\prime)(\lambda) + \Phi^-_r(d^\prime)(\lambda)
\leq d^\prime(\lambda) + 2^{-r}$.  This holds because
$\Phi^+_r(d^\prime)(\lambda) < 2^{-r}$ and
$\Phi^-_r(d^\prime)(\lambda) = d^\prime(\lambda)$.
\end{enumerate}
We have now seen that $\Phi$ is a $\Delta$-$\nu$-measurement of $X$.
Since $\Phi^+_r(\one)(\lambda) \leq 2^{-r}$ for each $r \in \N$, it
follows that $\nu_\Delta(X) = \Phi^+_\infty = 0$, confirming (1).
\end{proofof}

Analogous characterizations of the sets of $\nu$-measure 0 in
$R(\Delta)$ follow immediately from Theorem~\ref{Z.7} via
Lemma~\ref{Z.6}.

Note that conditions (2) through (5) in Theorem~\ref{Z.7} do not
involve type-2 functionals.  It is these characterizations of measure
$0$ sets in terms of martingales (type-1 objects) that have been used
in resource-bounded measure to date.

\section*{Acknowledgments} I thank Alekos Kechris and Yaser
Abu-Mostafa for their hospitality during several visits at Caltech,
where much of this research was performed.  I thank Dexter Kozen and
Juris Hartmanis for their hospitality at Cornell, where much of the
writing took place.  I thank Klaus Weihrauch, Jim Royer, and Jack Dai
for pointing out errors in an earlier draft of this paper.  For
helpful discussions over the past few years, I thank many colleagues,
including Steve Fenner, David Juedes, Steve Kautz, Jim Lathrop, Elvira
Mayordomo, Ken Regan, Jim Royer, Giora Slutzki, Martin Strauss, and
Bas Terwijn.  I am especially grateful to my late friend Ron Book for
his constant encouragement of this project, and I dedicate this paper
to his memory.

\bibliography{main,random,rbm}
\bibliographystyle{plain}
%
%
%
%
%
%

\end{document}